\documentclass[9pt,twocolumn,twoside,hidelinks]{pnas-new}

\usepackage{siunitx}
\usepackage{microtype}

\usepackage{xfp}

\usepackage{latexsym}      
\usepackage{dcolumn}
\usepackage{hyperref}
\hypersetup{
 colorlinks=false,
 citecolor=false,
 linkcolor=false,
 urlcolor=false,
 pdfpagemode=UseNone,
 pdfstartview=FitH}

\usepackage{scalerel,stackengine,amsmath}

\usepackage[eulergreek]{sansmath}
\usepackage{ upgreek }

\usepackage{filecontents}

\usepackage{booktabs}
\usepackage{dcolumn}
\newcolumntype{d}[1]{D{.}{.}{#1}}

\usepackage{bm}
\renewcommand{\vec}[1]{\text{\bfseries #1}}

\usepackage{xcolor}
\usepackage{dcolumn}

\usepackage{tikz}
\usetikzlibrary{math}
\usetikzlibrary{positioning}
\usetikzlibrary{shapes.geometric, arrows}
\usetikzlibrary{arrows.meta}
\usetikzlibrary{positioning}
\usetikzlibrary{shapes,arrows}
\usetikzlibrary{intersections}
\usepackage{pgfplots}
\pgfplotsset{
tick label style = {font=\sansmath\sffamily},
every axis label/.append style = {font=\sansmath\sffamily}
}

\usetikzlibrary{arrows}
\pgfplotsset{compat=newest}
\usepgfplotslibrary{fillbetween}
\usetikzlibrary{calc}
\def\centerarc[#1](#2)(#3:#4:#5)
{ \draw[#1] ($(#2)+({#5*cos(#3)},{#5*sin(#3)})$) arc (#3:#4:#5); }

\pgfplotscreateplotcyclelist{usualsuspectinner}{
{mark=*,c1},
{mark=triangle*,c4},
{mark=diamond*,gold},
{mark=halfcircle*,c11},
{mark=square*,c21}, 
{mark=halfsquare left*,c23},
{mark=pentagon*,c28},
      }
\pgfplotscreateplotcyclelist{usualsuspectouter}{
{mark=o,c1},
{mark=triangle,c4},
{mark=diamond,gold},
{mark=halfcircle,c11},
{mark=square,c21}, 
{mark=halfsquare left,c23},
{mark=pentagon,c28},
      }

\pgfplotsset{
    every first x axis line/.style={},
    every first y axis line/.style={},
    every first z axis line/.style={},
    every second x axis line/.style={},
    every second y axis line/.style={},
    every second z axis line/.style={},
    first x axis line style/.style={/pgfplots/every first x axis line/.append style={#1}},
    first y axis line style/.style={/pgfplots/every first y axis line/.append style={#1}},
    first z axis line style/.style={/pgfplots/every first z axis line/.append style={#1}},
    second x axis line style/.style={/pgfplots/every second x axis line/.append style={#1}},
    second y axis line style/.style={/pgfplots/every second y axis line/.append style={#1}},
    second z axis line style/.style={/pgfplots/every second z axis line/.append style={#1}}
}

\makeatletter
\def\pgfplots@drawaxis@outerlines@separate@onorientedsurf#1#2{%
    \if2\csname pgfplots@#1axislinesnum\endcsname
    \else
    \scope[/pgfplots/every outer #1 axis line,
        #1discont,decoration={pre length=\csname #1disstart\endcsname, post length=\csname #1disend\endcsname}]
        \pgfplots@ifaxisline@B@onorientedsurf@should@be@drawn{0}{%
            \draw [/pgfplots/every first #1 axis line] decorate {
                \pgfextra
                \pgfplotspointonorientedsurfaceabsetupfor{#2}{#1}{\pgfplotspointonorientedsurfaceN}%
                \pgfplots@drawgridlines@onorientedsurf@fromto{\csname pgfplots@#2min\endcsname}%
                \endpgfextra 
                };
        }{}%
        \pgfplots@ifaxisline@B@onorientedsurf@should@be@drawn{1}{%
            \draw [/pgfplots/every second #1 axis line] decorate {
                \pgfextra
                \pgfplotspointonorientedsurfaceabsetupfor{#2}{#1}{\pgfplotspointonorientedsurfaceN}%
                \pgfplots@drawgridlines@onorientedsurf@fromto{\csname pgfplots@#2max\endcsname}%
                \endpgfextra 
                };
        }{}%
    \endscope
    \fi
}%
\makeatother

\usepackage{csvsimple}
\usepackage{changepage}
\usepackage[tbtags]{mathtools}
\usepackage{siunitx}[=v2]
\usepackage{csquotes}
\usepackage{enumerate}
\usepackage{enumitem}
\usepackage{float}
\usepackage[label font=bf]{subfig}
\usepackage{filecontents}

\usepackage[
	nonumberlist, 
	acronym, 
	nomain,
	nopostdot,
]{glossaries}

\usepackage{color, colortbl}

\usepackage[utf8]{inputenc}
\usepackage{subfig}
\usepackage{csvsimple}
\usepackage{tikz}
\usepackage{nicefrac}

\usepackage{changepage}
\usepackage{color, colortbl}

\usepackage{authblk}
\usepackage{cleveref}
\usepackage{siunitx}

\usepackage{blindtext}
\usepackage{amssymb}

\let\vec\mathbf
\usepackage{nicefrac}

\usepackage{multirow}
\usepackage{tabularx}
\usepackage{booktabs}
\usepackage{array}
\usepackage{longtable}
\newcolumntype{L}[1]{>{\raggedright\let\newline\\\arraybackslash\hspace{0pt}}m{#1}}
\newcolumntype{C}[1]{>{\centering\let\newline\\\arraybackslash\hspace{0pt}}m{#1}}
\newcolumntype{R}[1]{>{\raggedleft\let\newline\\\arraybackslash\hspace{0pt}}m{#1}}

\usepackage{filecontents}

\usepackage[normalem]{ulem}

\usepgfplotslibrary{external} 

\newacronym{3d}{3D}{three dimensional}
\newacronym{am}{AM}{additive manufacturing}
\newacronym{dem}{DEM}{discrete element method}
\newacronym{md}{MD}{molecular dynamics}
\newacronym{dc}{DC}{direct-current}
\newacronym[plural=PFCs,firstplural=parabolic flight campaigns (PFCs)]{pfc}{PFC}{Parabolic Flight Campaign}
\newacronym{fft}{FFT}{Fast Fourrier Transform}
\newacronym{cad}{CAD}{Computer Assisted Design}
\newacronym{ptfe}{PTFE}{polytetrafluoroethylene}
\newacronym{ps}{PS}{polystyrene}
\newacronym{nasa}{NASA}{National Aeronautics and Space Administration}
\newacronym{esamm}{ESAMM}{Extended Structure Additive Manufacturing Machine}
\newacronym{amf}{AMF}{Additive Manufacturing Facility}
\newacronym{us}{US}{United States}
\newacronym{usa}{USA}{United States of America}
\newacronym{bmgs}{BMGs}{Bulk Metallic Glasses}
\newacronym{esa}{ESA}{European Space Agency}
\newacronym{dlr}{DLR}{German Aerospace Center, abbreviated from German \textit{Deutsches Zentrum f\"{u} Luft- und Raumfahrt e.V.}}
\newacronym{liggghts}{LIGGGHTS}{\acrshort{lammps} Improved for General Granular and Granular Heat Transfer Simulations}
\newacronym{rpm}{rpm}{revolutions per minute}
\newacronym{rise}{RISE}{Research Internships in Science and Engineering}
\newacronym{daad}{DAAD}{German Academic Exchange Service, abbreviated from German \textit{Deutscher Akademischer Austauschdienst}}
\newacronym{fsm}{FSM}{finite-state machine}
\newacronym{ir}{IR}{infrared}
\newacronym{pcbs}{PCBs}{Printed Circuit Boards}
\newacronym{pcb}{PCB}{Printed Circuit Board}
\newacronym{mcr}{MCR}{modular compact rheometer}
\newacronym{sff}{SFF}{Solid Freeform Fabrication}
\newacronym{uv}{UV}{ultraviolet}
\newacronym{abs}{ABS}{acrylonitrile butadiene styrene}
\newacronym{hpde}{HPDE}{high density polyethylene}
\newacronym{pei}{PEI}{polyetherimide}
\newacronym{bff}{BFF}{BioFabrication Facility}
\newacronym{lens}{LENS}{Laser Engineered Net Shaping}
\newacronym{cnc}{CNC}{Computer Numerical Control}
\newacronym{ebf3}{EBF$^3$}{Electron Beam Free-Form Fabrication}
\newacronym{leo}{LEO}{Low Earth Orbit}
\newacronym{pc}{PC}{polycarbonate}
\newacronym{crissp}{CRISSP}{Customisable Recyclable International Space Station Packaging}
\newacronym{Athena}{Athena}{Advanced Telescope for High-ENergy Astrophysics}
\newacronym{lbm}{LBM}{Laser Beam Melting}
\newacronym{bam}{BAM}{Federal Institute for Materials Research and Testing, abbreviated from German \textit{Bundesanstalt f\"{u}r Materialforschung und-pr\"{u}fung}}
\newacronym{pbf}{PBF}{powder bed fusion}
\newacronym{eb}{EB}{Electron Beam}
\newacronym{2d}{2D}{two dimensional}
\newacronym{ft4}{FT4}{Freeman Technology 4 Powder Rheometer}
\newacronym{dsc}{DSC}{Differential Scanning Calorimetry}
\newacronym{pmma}{PMMA}{polymethylmethacrylate}
\newacronym{1g}{$1g$}{gravity on-ground}
\newacronym{mug}{$\mu g$}{microgravity}
\newacronym{bcm}{BCM}{Box Counting Method}
\newacronym{mct}{MCT}{mode coupling theory}
\newacronym{gmct}{g-MCT}{granular mode-coupling theory}
\newacronym{itt}{ITT}{integration through transients}
\newacronym{mfc}{MFC}{Mass Flow Controller}
\newacronym{ct}{CT}{computed tomography}
\newacronym{cv}{CV}{curriculum vitae}
\newacronym{pi}{PI}{principal investigator}
\newacronym{osp}{OSP}{orthogonal superimposed perturbation}
\newacronym{fps}{fps}{frames per second}
\newacronym{pdf}{pdf}{probability density function}
\newacronym{al}{Al}{aluminium}
\newacronym{ss}{\textit{SS}}{\textit{Smooth Surface}}
\newacronym{rs}{\textit{RS}}{\textit{Rough Surface}}
\newacronym{isru}{ISRU}{\textit{in-situ} resource utiliszation}
\newacronym{rcp}{rcp}{random close packing}
\newacronym{rlp}{rlp}{random loose packing}
\newacronym{gitt}{g-ITT}{granular integration through transient} 
\newacronym{si}{SI}{Supplementary Information}

\definecolor{c1}{rgb}{0.7068574918274737, 0.11027871818526241, 0.2747061222663145}%
\definecolor{c2}{rgb}{0.6780437401750237, 0.1857033496010309, 0.10475664313058389}%
\definecolor{c3}{rgb}{0.5819819292481591, 0.28135510723834917, 0.10365807489974799}%
\definecolor{c4}{rgb}{0.5234892407222805, 0.3183040932830017, 0.10308831855626377}%
\definecolor{c5}{rgb}{0.4801916866157751, 0.3399605081271155, 0.10271364194308724}%
\definecolor{c6}{rgb}{0.44359832760663015, 0.3553988979685888, 0.10242748858902176}%
\definecolor{c7}{rgb}{0.40913672397205286, 0.36794412723413256, 0.10218299803254591}%
\definecolor{c8}{rgb}{0.37320211354661986, 0.37925108836394167, 0.10195327596798023}%
\definecolor{c9}{rgb}{0.33135856975649947, 0.3904277497779026, 0.10171733001337946}%
\definecolor{c10}{rgb}{0.2751398809612443, 0.40252912490464393, 0.10145175401497963}%
\definecolor{c11}{rgb}{0.17705545280426335, 0.4169982861326329, 0.10112016988851782}%
\definecolor{c12}{rgb}{0.10370113411519366, 0.41987331701008007, 0.20784080256789766}%
\definecolor{c13}{rgb}{0.10672620229256541, 0.415683426104876, 0.2824387977867041}%
\definecolor{c14}{rgb}{0.10896052083074859, 0.412482459879469, 0.3262929409150867}%
\definecolor{c15}{rgb}{0.11082049183583692, 0.4097465781098524, 0.3585658834499722}%
\definecolor{c16}{rgb}{0.1125289398766243, 0.40717495121262287, 0.38576967807315987}%
\definecolor{c17}{rgb}{0.11424591411818866, 0.4045324469356048, 0.41127417284765605}%
\definecolor{c18}{rgb}{0.11613475317261168, 0.4015563987219263, 0.4376207984793854}%
\definecolor{c19}{rgb}{0.11843118069754927, 0.3978377246048391, 0.46769849466738594}%
\definecolor{c20}{rgb}{0.1215889511379443, 0.3925371130865723, 0.5062751486798138}%
\definecolor{c21}{rgb}{0.12675667510032929, 0.38336665224260497, 0.564172135389151}%
\definecolor{c22}{rgb}{0.13823697094516182, 0.36050804998003744, 0.6775165660716256}%
\definecolor{c23}{rgb}{0.2929229731919379, 0.2517707788576924, 0.9106622939003164}%
\definecolor{c24}{rgb}{0.5044517963791875, 0.15823641315504755, 0.8374462940274711}%
\definecolor{c25}{rgb}{0.5743505600217842, 0.14563681653078617, 0.7277475942695497}%
\definecolor{c26}{rgb}{0.6114384687170589, 0.13753415571759905, 0.6502354669913063}%
\definecolor{c27}{rgb}{0.6363457404298491, 0.13141740987926892, 0.586320806947322}%
\definecolor{c28}{rgb}{0.6557132733878439, 0.12622710616005786, 0.5268534496956088}%
\definecolor{c29}{rgb}{0.6725336221941806, 0.12137095721249477, 0.4649001891758376}%
\definecolor{c30}{rgb}{0.688593421429631, 0.11639483989072555, 0.39157406584043325}%
\definecolor{brickred}{rgb}{0.8, 0.25, 0.33}%
\definecolor{darkorange}{rgb}{1.0, 0.55, 0.0}%
\definecolor{persiangreen}{rgb}{0.0, 0.65, 0.58}%
\definecolor{persianindigo}{rgb}{0.2, 0.07, 0.48}%
\definecolor{cadet}{rgb}{0.33, 0.41, 0.47}%
\definecolor{turquoisegreen}{rgb}{0.63, 0.84, 0.71}%
\definecolor{sandybrown}{rgb}{0.96, 0.64, 0.38}%
\definecolor{blueblue}{rgb}{0.0, 0.2, 0.6}%
\definecolor{ballblue}{rgb}{0.13, 0.67, 0.8}%
\definecolor{greengreen}{rgb}{0.0, 0.5, 0.0}%
\definecolor{amber}{rgb}{1.0, 0.75, 0.0}%
\definecolor{goldenrod}{rgb}{0.85, 0.65, 0.13}%
\definecolor{gold}{rgb}{0.86, 0.71, 0.23}%
\definecolor{tiffanyblue}{rgb}{0.04, 0.73, 0.71}%
\definecolor{bittersweet}{rgb}{1.0, 0.44, 0.37}%

\definecolor{purpleheart}{rgb}{0.41, 0.21, 0.61}%

\definecolor{ca}{RGB}{127,59,8}%
\definecolor{cb}{RGB}{179,88,6}%
\definecolor{cc}{RGB}{224,130,20}%
\definecolor{cd}{RGB}{253,184,99}%
\definecolor{ce}{RGB}{254,224,182}
\definecolor{cf}{RGB}{216,218,235}%
\definecolor{cg}{RGB}{178,171,210}%
\definecolor{ch}{RGB}{128,115,172}%
\definecolor{ci}{RGB}{84,39,136}%
\definecolor{cj}{RGB}{45,0,75}%

\def\Qtouconvwithcylinder{0.010306294}

\def\PhiQzerosix{0.5183}
\def\PhiQzeroseven{0.5149}
\def\PhiQzeroeight{0.5114}
\def\PhiQone{0.5089}
\def\PhiQonefive{0.5070}
\def\PhiQtwo{0.5065}
\def\PhiQthree{0.5062}

\def\phiINF{0.506}
\def\phiRLP{0.54}

\def\uzero{0.0021}
\def\ufp{0.0034}
\def\exposant{2}

\def\etaequation{2.6}
\def\phig{0.519}

\DeclareMathOperator{\Wi}{Wi}
\DeclareMathOperator{\Pe}{Pe}
\DeclareMathOperator{\St}{St}

\DeclareRobustCommand{\orderof}{\ensuremath{\mathcal{O}}}

\templatetype{pnasresearcharticle} 

\title{The manifold rheology of fluidized granular media}

\author[a,b,1]{Olfa D'Angelo}
\author[c]{Abhishek Shetty} 
\author[b,d]{Matthias Sperl}
\author[d,b]{W. Till Kranz}

\affil[a]{Institute for Multiscale Simulation, Erlangen-N\"{u}rnberg Universit\"{a}t, Cauerstra\ss{}e 3, 91058 Erlangen, Germany}
\affil[b]{Institute of Materials Physics in Space, German Aerospace Center (DLR), Linder H\"{o}he, 51170 Cologne, Germany}
\affil[c]{Rheology Department, Anton Paar USA, Inc., 10215 Timber
  Ridge Drive, Ashland, VA 23005, United States of America}
\affil[d]{Institute for Theoretical Physics, University of Cologne, 50937 Cologne, Germany}

\leadauthor{D'Angelo} 

\significancestatement{From industrial reactors to geophysical flows,
fluidized granular media under shear are widespread, 
yet poorly understood from a fundamental point of view. 
Combining rheological measurements on air-fluidized glass beads 
spanning eight orders of magnitude in shear rate, 
with a glassy dynamics theory based on first-principles,
we confirm and explain rheological regimes in agitated granular beds:
Newtonian, shear thinning, and Bagnolidan shear thickening regimes.
Such carefully measured rheology provides essential ground-truth 
for future studies, notably to predict pyroclastic flows and avalanches.
It is also a step towards a universal theoretical description of granular flows.} 

\authorcontributions{
M.S.: conceptualization; project administration; validation; resources; writing - review \& editing.
W. T. Kranz: conceptualization; formal analysis; methodology; validation; supervision; writing - original draft.
O.D'A.: conceptualization; data curation; investigation; formal analysis; methodology; project administration; visualization; writing - original draft.
A.S.: resources; data curation; writing - review \& editing.
}  
\authordeclaration{The authors declare no conflict of interest.}

\correspondingauthor{\textsuperscript{1}To whom correspondence should be addressed. E-mail: olfa.dangelo@fau.de}


\begin{abstract}
Fluidized granular media 
have a rich rheology:
measuring shear stress $\bm\sigma$ as a function of shear rate $\bm\dot\gamma$,
they exhibit Newtonian behavior $\bm{\sigma\sim\dot\gamma}$ for low densities and shear rates, 
develop a yield stress 
for intermediate shear rates and densities approaching the granular glass transition,
and finally, cross over to shear-thickening Bagnold scaling,
$\bm{\sigma\sim{\dot\gamma^2}}$. 
This wealth of flow-behaviors makes fluidized beds a fascinating material,
but also one that is challenging to encompass into a global theory,
despite its relevance for optimizing industrial processes and predicting natural hazards.
We provide careful measurements spanning eight orders of magnitude in shear rate,
and show that all these rheological regimes can be described
qualitatively and quantitatively 
using the \acrlong{gitt} formalism, a theory for glassy dynamics under
shear adapted to granular fluids.
\end{abstract}

\dates{This manuscript was compiled on \today}

\begin{document}

\maketitle
\thispagestyle{firststyle}
\ifthenelse{\boolean{shortarticle}}{\ifthenelse{\boolean{singlecolumn}}{\abscontentformatted}{\abscontent}}{}

\dropcap{F}luidization counteracts the dissipative nature of granular
materials by a constant energy input~\cite{Geldart1973, FluidizationEngineeringBook1991, Zenz1997, Horio2017}.
Transforming static granular solids into dynamic granular fluids, fluidization maintains a
non-zero mean kinetic energy per particle, i.e., a finite
granular temperature $T$~\cite{Goldhirsch2008}.
The macroscopic solid particles that collectively
constitute the granular medium may be agitated by a number of external
forces, including an interstitial fluid pressure in gas-fluidized
media~\cite{Bakhtiyarov1999, Roche2004, Kottlan2018, Breard2022}, shaking in
vibro-fluidization~\cite{Dijksman2011,Zhang2018,Gnoli2016}, pressure
waves in acoustic fluidization~\cite{Melosh1996,Conrad2013}.

In most natural and human-made processes involving granular fluids,
fluidization and shear happen simultaneously.
Typical examples in geophysical flows~\cite{Melosh1996, Wilson1984,
  Roche2004, Kelfoun2009, Breard2022} 
are avalanches, land slides, pyroclastic flows --
all natural hazards threatening human lives. 
Two-phase flows of a
non-Brownian solid phase of macroscopic particles in a molecular fluid
are also a common industrial system, 
e.g.~in fluid-particles mixing~\cite{Eames2005}, 
fluidized bed reactors~\cite{Werther2007, Menendez2019}, 
or pneumatic conveyors~\cite{Dhurandhar2018}.

Despite the critical importance of fluidized granular media under
shear for process optimization and risk mitigation, studies of their
rheology remain sparse and produce seemingly conflicting results. 
Some studies find Newtonian rheology
\cite{Matheson1949,Hagyard1966,Maeno1980,Koos2012}, some shear
thinning \cite{Bouwhuis1961,Tardos1998}, and some Bagnold rheology
\cite{Hanes1985}. Other studies found crossovers between these
\cite{Kottlan2018,Schugerl1961,Hobbel1985,Gottschalk1986,Mishra2020,Young2021}. 
We will show below that, in fact, the fluidized bed rheology comprises
all three rheological behaviors: Newtonian, shear thinning, and
Bagnoldian. For the Bagnold regime \cite{Bagnold1954,Lois2005} alone,
the empirical $\mu(\mathcal I)$-rheology \cite{Midi2004} has been very
successful.
However, a succinct and comprehensive constitutive
equation describing agitated granular fluids -- stress tensor $\Sigma$
as a function of shear rate $\dot\gamma$ -- remains hitherto missing.

We carefully measure the steady-state rheology of gas-fluidized glass
beads, treating the fluidized bed as an effectively homogeneous
material, and measuring the macroscopic mean shear stress $\sigma$ as
a function of shear rate $\dot{\gamma}$, in a wide-gap Taylor-Couette
shear cell \cite{Francia2021}.
By spanning eight orders of magnitude in shear rates, we delineate a
diversity of rheological regimes.
Using a glassy dynamics theory, the \gls{gitt} formalism as recently
proposed by some of us as a constitutive model, derived from first
principles~\cite{Kranz2018, Kranz2020}, we fit the experimental flow
curves. Using a minimal number of physically meaningful fit
parameters, all rheological regimes are encompassed in the theoretical
prediction. Finding that the fit parameters are consistent among
different flow curves, we thereby also validate the \gls{gitt}
constitutive relation.

\section*{Gas-fluidized granular media under shear}

\begin{figure*}[bt!]
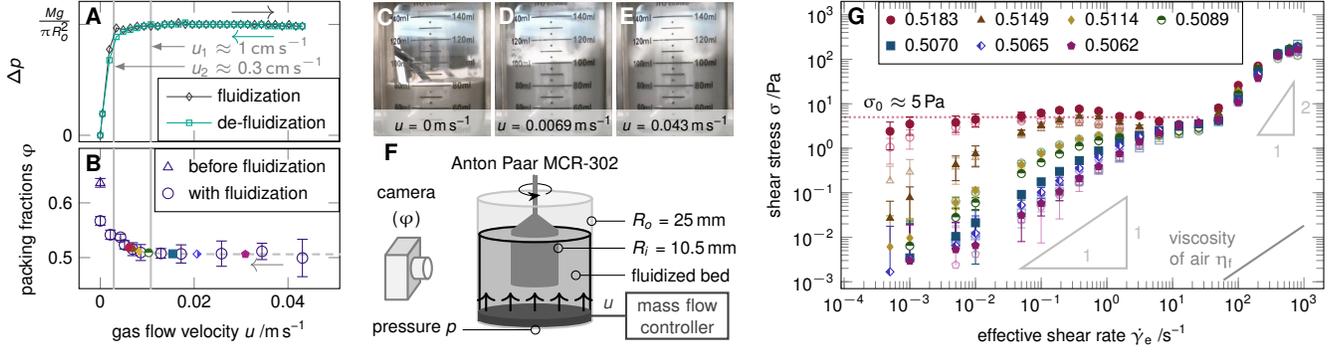

  \centering
  \begin{tikzpicture}
\sisetup{detect-all}
\normalfont\sffamily
\sansmath\sffamily
\footnotesize
 
\begin{scope}[]
\input{pressure-drop.tex}
\end{scope}

\begin{scope}[xshift=0.162\textwidth]
\input{fluidized-bed-pictures}
\input{rheometer-schematic}
\end{scope}

\begin{scope}[xshift=0.57\textwidth,]
\input{stress_flowcurves.tex}
\end{scope}

\end{tikzpicture} 
  \caption{\label{fig:experiment}
\sisetup{detect-all}\normalfont\sansmath\sffamily
    Fluidization and flow curves of air-fluidized glass beads.
(A)~Air pressure drop across the powder bed,
    $\mathsf{\Delta} p$, due to the flow resistance of the granular
    bed, as a function of superficial fluidization velocity $u$, for
    decreasing (green squares) followed by increasing (gray diamonds)
    air flow velocity $u$.  (B)~Global packing fraction $\varphi$ as a
    function of air flow velocity $u$ (colored symbols indicate the
    flow velocities selected for further investigation, same color
    coding as in Fig.~\ref{fig:experiment}G -- see legend there for
    packing fraction values). Error bars correspond to the volume fluctuation due to bubbling.
	Vertical lines in (A, B) indicate the
    lower and upper bounds of the fluidization range (respectively
    $u_2$ and $u_1$). (C, D, E)~Pictures of the powder bed at
    increasing air flow velocity, from left to right:
    $u=$~\SI{0}{\centi\meter\per\s} (before fluidization),
    $u=$~\SI{0.69}{\centi\meter\per\s} and
    $u=$~\SI{4.3}{\centi\meter\per\s}. (F)~Schematic of the wide-gap
    Taylor-Couette rheometery setup: Anton Paar MCR~302 with
    commercial powder flow cell and profiled cylinder (see Methods for
    details), and camera recording the global powder bed volume.
    (G)~Macroscopic stress $\sigma$ as a function of effective shear rate $\dot\gamma_\text{e}$ 
    for the air flow velocities indicated in~(B).
    Each mark shape and color represent one air flow velocity $u$; equivalent $\varphi$ are given as legend.
    Filled marks represent the upward sweep, empty marks represent downward sweep in $\dot{\gamma}_\text{e}$. Error bars are the standard deviation on all data points measured in steady state.
    The shear stress due to air alone is indicated by the gray line at the bottom right of the graph.
    Triangles at low and high shear rates indicate the slopes of
    Newtonian and Bagnold behavior, respectively. Dotted red
    horizontal line indicates the yield stress of the partially
    fluidized bed at $\sigma_0 \approx$~\SI{5}{\pascal}. }
\end{figure*}

\subsection*{Fluidized bed}
\label{sec:fluidization}

Gas-fluidized granular fluids are a challenging material for a number
of reasons. They are a genuine two-phase material, composed of the
macroscopic solid particles that constitute the granular phase and the
molecular fluid surrounding them
\cite{Lemaitre2009,Guazzelli2018,Wang2020}. Stokes numbers are
$\orderof(10)$, such that the particles are neither completely
over-damped and enslaved by the molecular fluid~\cite{Guazzelli2018}, 
nor can the molecular fluid's
viscosity, $\eta_\text{f}$, be neglected to treat the material as a
granular gas of particles only. 
As the particles are
macroscopic, they are athermal, i.e.~their kinetic energy is not in
equilibrium with the thermal energy of the molecular fluid.
Nevertheless, as the gas flow maintains the grains' motion, a finite
mean kinetic energy per particle can be associated with any granular
fluid, i.e., a finite granular temperature $T > 0$, unrelated to the
thermodynamic temperature. The energy injected by fluidization is
constantly dissipated in particle collisions, making granular
fluids by nature out of equilibrium.

We focus on a fluidized bed of glass beads of mean diameter
$d = \SI{69}{\micro\metre}$ (Geldart group A~\cite{Geldart1973}, see
Methods for details and Refs.~\cite{LaMarche2016,Mishra2020} for further
characterization), and vary the fluidization velocity $u$. 
To characterize the fluidization state, we monitor the pressure drop 
$\Delta p$ across the particle bed (Fig.~\ref{fig:experiment}A). 
While we find the bed to be fully fluidized at $u > u_1$
($Mg = \Delta p\pi R_0^2$, 
with sample mass $M=$~\SI{125}{\g}, 
the gravitational acceleration $g$, and a container of radius $R_0$),
it is already
partially fluidized at $u_1 > u > u_2$, as fluidization occurs across
a fluidization range~\cite{Grace2008}. We observe only a small
pressure overshoot in the fluidization \textit{versus} defluidization
curve around $u_1$, which indicates minimal cohesion between
particles~\cite{Valverde1998}.

By recording the time-averaged mean bed height
(Figs.~\ref{fig:experiment}C--E), we determine the global packing
fraction $\varphi(u)$ corresponding to a given fluidization velocity
$u$ (see Fig.~\ref{fig:experiment}B and Methods). Starting in the
partially fluidized regime and reaching deep into the fully fluidized
state, we select seven representative fluidization states -- or
packing densities -- for further investigation. Note that we report
the packing fractions to a much higher precision than is warranted by
the measurement's accuracy to reflect the monotonicity of $\varphi(u)$
\cite{jackson00}. The superficial fluidization velocity $u$ remains the
primary control parameter. At the higher fluidization velocities, we
observe significant bubbling \cite{Geldart1973,jackson00}. However, by
time-averaging, we treat the fluidized bed, including bubbles, as an
effectively homogeneous material.

\subsection*{Rheometry}\label{subsec:rheometry}

We measure the steady-state macroscopic shear stress
$\sigma(\dot{\gamma}_\mathrm{e})$ of air-fluidized glass beads, as a
function of effective shear rate $\dot\gamma_\mathrm{e}$.  
An Anton Paar MCR~302 rheometer
with Taylor-Couette geometry
is used in shear-rate controlled mode
(see Fig.~\ref{fig:experiment}F and Methods for technical details).
Fig.~\ref{fig:experiment}G shows the resulting flow curves (see
Methods for full experimental procedure); $\sigma$ is time-averaged per
$\dot\gamma_\emph{e}$, and for every packing fraction, we measured the
curves both in an up-sweep and down-sweep in shear rate. 
We call effective shear rate the shear rate as calculated
for a purely Newtonian fluid experiencing uniform shear across the
entire gap, i.e., $\dot\gamma_\mathrm{e} := 2\pi K_\text{N} U$, linearly
related to the rotation rate of the inner cylinder $U$ 
by the shear constant $K_\text{N}$ (see Methods
for details). For preliminary results with a similar setup see
Refs.~\cite{Kottlan2018, Kranz2022}.

We see that irrespective of the bubbling instability, packing
fractions are typically high, $\varphi \gtrsim 0.5$, such that a
fluidized bed cannot be regarded as a dilute suspension, where the
suspended particles merely provide a correction to the molecular
fluid's viscosity $\eta_\text{f}$. On the contrary, the shear stress
is dominated by the particles and orders of magnitude larger than the
stress $\eta_\mathrm{f}\dot\gamma_\text{e}$ exerted by air alone.

\section*{Rheological Regimes}
\label{sec:rheological-regimes}

\begin{figure*}[bt!]
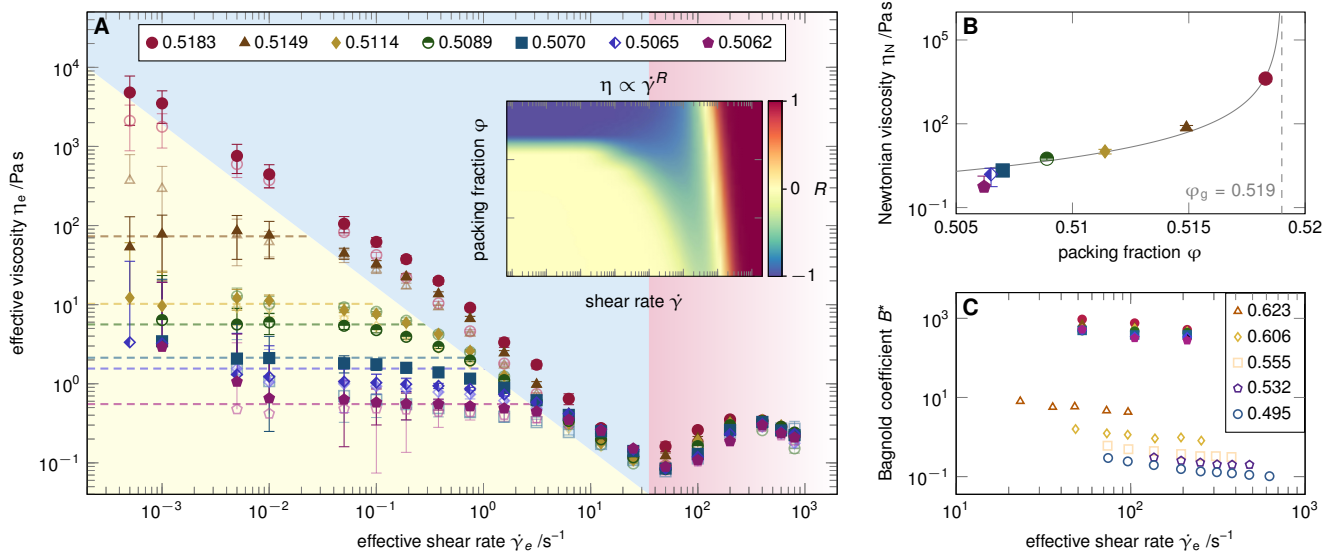

  \centering
  \begin{tikzpicture}
\sisetup{detect-all}
\normalfont\sffamily
\sansmath\sffamily
\footnotesize
 
\begin{scope}[]
\input{viscosity-flowcurve.tex}
\end{scope}

\def\shiftiny{0.21}
\def\sizex{0.35}
\def\sizey{0.24}
\begin{scope}[xshift=0.65\textwidth,]
\input{viscosityVSphi}
\input{BagnoldCoefVSphi}
\end{scope}

\end{tikzpicture} 
  \caption{\sisetup{detect-all}\normalfont\sansmath\sffamily
Experimental rheology of an air-fluidized granular bed.
    (A)~Apparent viscosity,
    $\eta_\text{e}(\dot\gamma_\text{e}) = \sigma/\dot\gamma_\text{e}$, as a
    function of effective shear rate $\dot\gamma_\text{e}$ for
    different packing fractions $\varphi$ (see legend). Error bars
    are the standard deviation on all data points measured in steady
    state.  The different rheological regimes are indicated by the
    shaded background. 
Dashed lines indicate the Newtonian viscosities $\eta_\mathrm{N}$.
The inset represents the rheological regimes
    predicted by \acrfull{gitt} (see body of text for details), where
    the color map corresponds to values of $R$ in
    $\eta \propto \dot{\gamma}^R$. 
(B)~Newtonian
    viscosities $\eta_\mathrm{N}(\varphi)$ as a function of
    packing fraction $\varphi$. Error bars are standard deviation on
    all values of $\eta_\text{e}(\dot{\gamma})$ within the
    Newtonian regime. A fit with the law
    $\eta_\mathrm{N}(\varphi) \propto (\varphi_g -
    \varphi)^{-\vartheta}$ predicts a divergence at the granular glass
    transition $\varphi_\mathrm{g} = \phig$ and $\vartheta = \etaequation $.  
(C)~Bagnold coefficients $B = \sigma/{\dot\gamma}^2$ (full symbols) as a
    function of the effective shear rate $\dot\gamma$ and for the different
    packing fractions (mark shape \& color code as in (A)). The lower sets of points
    (empty symbols) are the Bagnold coefficients extracted from
    Bagnold's historic measurements, see Ref.~\cite{Bagnold1954} (packing
    fractions indicated in legend).
}
  \label{fig:measurements}
\end{figure*}

We observe a complex rheology of the fluidized bed, with regions of
the parameter space that appear to be compatible with a Newtonian
rheology $\sigma\sim\dot\gamma_e$, parts that indicate Bagnold scaling
$\sigma\sim{\dot\gamma}_e^2$, and a shear thinning regime where
$\sigma$ appears almost constant. 
These rheological regimes become more blatant 
if we plot the shear rate dependent apparent viscosity
$\eta_\text{e}(\dot\gamma_\text{e}) \equiv \sigma /
\dot{\gamma}_\text{e}$ as in Fig.~\ref{fig:measurements}A.

\subsection*{Newtonian Regime}

At low shear rates and densities, the effective viscosity
$\eta_\text{e}(\dot\gamma_{\text{e}}\mid\varphi) \simeq
\eta_\text{N}(\varphi)$ assumes a shear rate independent value that
strongly depends on volume fraction, displaying Newtonian rheology
(yellow region in Fig.~\ref{fig:measurements}A).  In
Fig.~\ref{fig:measurements}B, we plot the viscosity in the Newtonian
regime as a function of packing fraction, $\eta_\text{N}(\varphi)$
(numerical values given in Tab.~\ref{tab:values}).

A fit to the following ansatz,
\begin{equation}
  \label{eq:1}
  \eta_\text{N}(\varphi) \propto (\varphi_\text{g} - \varphi)^{-\vartheta}, 
\end{equation}
suggests a viscosity diverging towards
$\varphi_\text{g} = \phig$, with exponent
$\vartheta\approx \etaequation$, where we interpret $\varphi_\text{g}$
as the critical packing fraction at the granular glass transition.
Note that the value of $\varphi_\text{g}$ (even within the
substantial systematic uncertainty of the average volume fraction
measurement of $\pm 4.6\%$) 
could be seen as surprisingly low, 
if Eq.~\textbf{\ref{eq:1}} were to be read as the Krieger-Dougherty
relation \cite{Krieger1959} of colloidal suspensions. 
Furthermore, interpreted as a close packing density, 
one would expect $\varphi_\text{g}$ close to
jamming, $\varphi_\text{J}\approx0.64$ 
(or slightly lower given surface friction and polydispersity~\cite{Song2008}). 
However, there is evidence
of a granular glass transition, 
i.e., a dynamically arrested, non-classical granular solid 
at $T>0$~\cite{abate+durian06,Kranz2010}.
Our analysis supports a kinetic granular glass transition~\cite{Seguin2016}, 
where \gls{gmct}~\cite{Kranz2010} predicts Eq.~\textbf{\ref{eq:1}} already
at a critical density $\varphi_\text{g}$ much lower than
$\varphi_\text{J}$. We will come back to the value of
$\varphi_\text{g}$ in the discussion.

\subsection*{Bagnold Regime}
At high shear rates
$\SI{50}{\per\s} \lesssim \dot\gamma_\text{e}
\lesssim\SI{350}{\per\s}$ (red region in
Fig.~\ref{fig:measurements}A), $\sigma \sim \dot{\gamma}^2 $,
exhibiting a shear thickening regime known as Bagnold
regime~\cite{Bagnold1954}.

As the material's flow profile varies with its rheology, the true shear rate
is no longer given by
$\dot\gamma_\mathrm{e}$, but related to the turning rate of the inner
cylinder, $U$, as $\dot\gamma = K_\text{B} 2\pi U$, with the strain constant
\begin{equation}
  \label{eq:KB}
  K_\text{B} = \frac{\delta}{\delta - 1},
\end{equation}
where $\delta = R_o / R_i$, the ratio of outer to inner radii of the
shear cell (full derivation is available as \gls{si}).

After the seminal measurements by Bagnold~\cite{Bagnold1954}, the
Bagnold regime and its emergence have
been studied repeatedly~\cite{Forterre2008, Madraki2020, Tapia2022},
although quantitative measurements of Bagnold coefficients are still
scarce.  We report in Fig.~\ref{fig:measurements}C the resulting
Bagnold coefficients $B = \sigma/{\dot\gamma}^2$, in dimensionless
form $B^* = Bd/m$ ($d$ and $m$ are respectively the average
diameter and mass of one particle).
Numerical values of $B(\varphi)$ are given in Tab.~\ref{tab:values}.

We can compare our results with Bagnold's original
measurements~\cite{Bagnold1954}, also plotted in
Fig.~\ref{fig:measurements}C.
The granular suspension in Ref.~\cite{Bagnold1954} are wax
balls of \SI{1.3}{\mm} diameter, suspended in water or a similar
density mixture of glycerin, alcohol and water. Glass beads being much
less dissipative than wax balls, our Bagnold
coefficients are orders of magnitude larger than
Bagnold's~\cite{Jenkins2007,Kranz2020}.

\begin{table}[tbhp]
\sisetup{detect-all}\normalfont\sffamily\sansmath\sffamily
\centering
\caption{\sisetup{detect-all}\sansmath\label{tab:values}
For every air flow velocities $\bm u $ studied: measured global packing fraction $ \bm\varphi$;
average viscosity in the Newtonian regimes $\bm \eta_\text{N}$ (the range of corresponding shear rates is given,
calculated for upward sweep in $\bm{\dot{\gamma}}_\text{e}$);
Bagnold coefficients $\bm B$, obtained by averaging $\bm{B}\bm{(\dot{\gamma}_\mathrm{e})}$ for all $\bm{\dot{\gamma}}_\mathrm{e}\bm{\in}$~[50,\,200]\,\si{\per\s}.
}
\begin{tabular}{S[round-mode=figures]cS[round-mode=figures,exponent-mode=input]cS[round-mode=figures]}
\toprule
  $u$~/\si{\m\per\second}	&	$\varphi$	&
  $\eta_\text{N}$~/\si{\pascal\second} &
  $\dot{\gamma}_\text{e}$ range for $\eta_\text{N}$~/\si{\per\s} &
  $B$~/\si{\milli\pascal\square\s}\\
\midrule
\fpeval{0.6*\Qtouconvwithcylinder}	& \PhiQzerosix &     4154.62	& $5 \cdot 10^{-4}$&	 4.62	\\
\fpeval{0.7*\Qtouconvwithcylinder}	&	\PhiQzeroseven 	&	 73.09 	&	$[5 \cdot 10^{-4}, \, 0.01]$	&	 3.75	\\
\fpeval{0.8*\Qtouconvwithcylinder}	&	\PhiQzeroeight		&	 10.23 	&	$[5 \cdot 10^{-4}, \, 0.1]$		&	 3.25	\\
\fpeval{1*\Qtouconvwithcylinder}	&	\PhiQone		&	 5.62 	&	$[1 \cdot 10^{-3}, \, 0.1]$		&	 2.92	\\
\fpeval{1.5*\Qtouconvwithcylinder}	&	\PhiQonefive		&	 2.13	&	$[1 \cdot 10^{-3}, \, 0.19]$	&	 2.61	\\
\fpeval{2*\Qtouconvwithcylinder}	&	\PhiQtwo		&	 1.55	&	$[5 \cdot 10^{-3}, \, 0.75]$	&	 2.46	\\
\fpeval{3*\Qtouconvwithcylinder}	&	\PhiQthree		&	 0.55		&	$[1 \cdot 10^{-2}, \, 3.13]$	&	 2.32	\\
\bottomrule
\end{tabular}
\end{table}

\subsection*{Shear Thinning Regime}

The shear thinning regime is indicated by a blue shading in
Fig.~\ref{fig:measurements}A and comprises an intermediate range of
shear rates that widens towards higher densities. From the
corresponding stress plateau in Fig.~\ref{fig:experiment}G for our highest density $\varphi = 0.5183$,
we infer an apparent yield stress $\sigma_0(\varphi_\text{g})\approx\SI{5}{\pascal}$ at $\varphi_\text{g} = \phig$.
Note that the emergence of a yield stress at
$\varphi_\mathrm{g}$ is another manifestation of the fluidized bed's
transition from a granular fluid into an granular glassy solid~\cite{abate+durian06,Kranz2010,Seguin2016}.

On closer inspection, Figs.~\ref{fig:experiment}G and
\ref{fig:measurements}A reveal that the flow curves are non-monotonic
in the shear thinning regime. Such a phenomenology has been shown to
derive from shear banding \cite{Spenley1996,Fielding2007} and is to be
expected given the combination of a continuously varying shear stress
in the Taylor-Couette geometry and a nearly shear rate independent
regime in the constitutive relation. While in the Newtonian and
Bagnold regimes, the true shear rate $\dot\gamma$ is linearly related
to the turning rate $U$ by the strain constants, $K_\mathrm{N}$ and
$K_\mathrm{B}$, respectively, the mapping $\dot\gamma(U)$ is
non-linear in the shear thinning, and in particular in the shear
banding regime.  To find the true shear rate, we numerically analyze
shear banding in the next section.

\subsection*{High Shear Rate Regime}
\label{sec:high-shear-rate}

At the highest shear rates, a second shear thinning regime seems to
establish, clearly visible in Fig.~\ref{fig:measurements}A for
$\dot{\gamma}_\text{e} \geq \SI{400}{\per\s}$. 
Such second viscosity decrease past the shear-thickening Bagnold regime 
was observed experimentally, albeit rarely, 
in dense non-Brownian solid-liquid suspensions~\cite{Stickel2005, Barnes1987}.
One possible explanation is that very high shear rates create a pressure gradient in the radial direction, resulting in a strong anisotropy, and in depletion along that direction~\cite{Barnes1987}.
To the best of our knowledge, 
it was never observed in dry granular media before.

We propose a similar explanation:
the strong azimuthal air flow 
due to high rotation speed of the corrugated cylinder might create
an air cushion between the
fluidized bed and the inner cylinder.
In other words, the density along the radial direction becomes strongly heterogeneous,
rendering our interpretation of the fluidized bed as a
continuous material inapplicable 
at such high $\dot{\gamma}_\text{e}$.

\begin{figure*}[bt!]
  \centering
  \begin{tikzpicture}
\sisetup{detect-all}
\normalfont\sffamily
\sansmath\sffamily
\footnotesize
 
\begin{scope}[]
\input{ShearBanding-fit-parameters}
\end{scope}

\begin{scope}[xshift=0.48\textwidth]
\input{ShearBanding.tex}
\end{scope}

\end{tikzpicture} 
  \caption{\sisetup{detect-all}\normalfont\sansmath\sffamily
    \label{fig:shearbanding} Shear banding analysis to determine the
    true shear rate.
(A)~The non-monotonic experimental flow curves
    (triangle symbols, here for $u = $~\num[round-mode = places, round-precision = 3]{\fpeval{\Qtouconvwithcylinder*0.7}}\,\si{\meter\per\s},
see Fig.~S2 in \gls{si} for all other fluidization speeds) can be fitted to two
    \gls{gitt} curves, $\sigma^\mathrm{i}(\dot\gamma_\mathrm{e})$ (inner, solid),
    and $\sigma^\mathrm{o}(\dot\gamma_\mathrm{e})$ (outer, dashed),
    respectively, outside of the negative slope, shear banding
    regime. (B)~True shear rate, $\dot\gamma(r)$, as a function of
    radial coordinate $r$ in the shear cell for three different
    effective shear rates $\dot\gamma_\text{e}$ indicated in (A).
The lowest (orange) and highest (violet) effective shear rates
    $\dot\gamma_\mathrm{e}$ are respectively Newtonian (following
    $\sigma^\mathrm{o}$) and Bagnoldian (following
    $\sigma^\mathrm{i}$) throughout, while at the intermediate
    $\dot\gamma_\mathrm{e}$ (red) the sheared fluid radially separates
    into a Bagnoldian part near the inner cylinder (following
    $\sigma^\mathrm{i}$) and a Newtonian part near the outer cylinder
    (following $\sigma^\mathrm{o}$) with a discontinuous step at
    $r=r^*$. (C)~Apparent ideal gas pressure $nT_0(u)$ and apparent
    collision frequency $\omega_0(u)$ in the unsheared fluidized bed
    as a function of fluidization velocity $u$ as a result of the
    \gls{gitt} fit. (D)~Distance from the granular glass transition
    $\Updelta \varphi_{\mathrm{th}}$ in the \gls{gitt} fit \textit{versus} the
    distance to the granular glass transition $\Updelta\varphi$ from
    the measurements. An effective coefficient of restitution
    $\varepsilon=0.8$ with a corresponding
    $\varphi_\mathrm{g}^\mathrm{MCT}(\varepsilon=0.8) = 0.528$ is assumed
    in (C,D). For other choices of $\varepsilon$ see Fig.~S9 in \gls{si}.
(E)~Shear
    rate $\dot\gamma$ at the inner cylinder (solid lines) and at the
    outer cylinder (dashed lines), as a function of the effective
    shear rate $\dot\gamma_\mathrm{e}$ for the packing fractions
    studied. The thick black lines indicate the analytical relations
    [Eqs.~\textbf{\ref{eq:KB},\ref{eq:4}}] in the Newtonian and
    Bagnold regime. Note the increasing difference between inside and
    outside in the shear banding regime. For color coded packing
    fractions in (D, E), refer to Fig.~\ref{fig:experiment}G.}
\end{figure*}

\section*{Flow regimes \& non-monotonicity: Theoretical analysis}
\label{sec:analys-shear-thinn}

The rheological regimes observed are
reminiscent of the rheological state
diagram predicted by \gls{gitt}~\cite{Kranz2018}, shown in the inset
of Fig.~\ref{fig:measurements}A, and suggests to use \gls{gitt} 
to analyze our data. In fact, we are going to calibrate the \gls{gitt}
constitutive relation with the measured data, taking shear banding
into account. This will, in particular, allow us to obtain the true
shear rate, also in the shear thinning regime;
by presenting the measured stress as a function of true shear rate,
we deduce the constitutive relation of the fluidized bed under shear.

\subsection*{Granular Integration Through Transients}
\label{sec:gran-integr-thro-1}

The \gls{gitt} constitutive relation
$\sigma(\dot\gamma/\omega_0|\Delta\varphi)/nT_0$ has been derived from
first principles for a randomly driven fluid of monodisperse inelastic
smooth hard spheres, characterized by a constant coefficient of
restitution $\varepsilon$ (see Methods for details). Its shape is
controlled by the distance
$\Delta\varphi_\mathrm{th} =
\varphi_\mathrm{g}^{\mathrm{MCT}}(\varepsilon) - \varphi$ to the granular
glass transition $\varphi_\mathrm{g}^{\mathrm{MCT}}(\varepsilon)$, as
determined by \gls{gmct}, which depends on the coefficient of
restitution $\epsilon$~\cite{Kranz2010}.  Moreover, the constitutive
relation is only predicted up to an overall stress scale, $nT_0$,
equivalent to the ideal gas pressure in the unsheared granular fluid,
and a rate scale, $\omega_0$, the collision frequency in the unsheared
fluid. The \gls{gitt} constitutive relation is therefore parametrized
by the triple ($nT_0$, $\omega_0$, $\Delta\varphi_\text{th}$).

Qualitatively, \gls{gitt} takes into account two key aspects of
fluidized beds: (i) a finite structural relaxation time $\tau$ that
diverges upon approaching the granular glass transition, and (ii) the
fact that the granular temperature is set by the power balance between
on one hand,
the energy injected by fluidization, $u\Delta p/h$ ($h$ being the bed
height) and shear heating, $\sigma\dot\gamma$, and on the other hand, 
the energy dissipated in particle collisions. 
As a result, \gls{gitt} predicts:
Newtonian rheology for small Weissenberg numbers
$\Wi := \dot\gamma\tau \ll 1$, i.e., small densities and shear rates;
shear thinning for $\Wi\gg1$; Bagnold scaling once shear heating
dominates over fluidization at large shear rates
\cite{Kranz2010}. This is corroborated by the rheological regimes
described above.

\begin{figure*}[bt!]
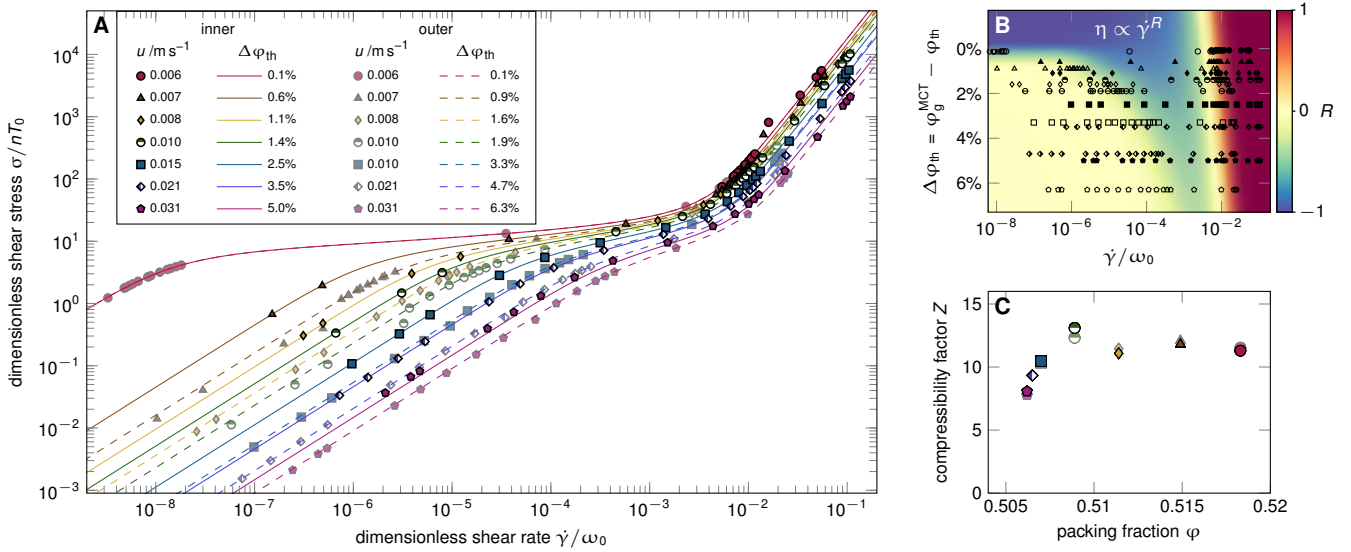

  \centering
\begin{tikzpicture}
\sisetup{detect-all}
\normalfont\sffamily
\sansmath\sffamily
\footnotesize
 
\begin{scope}[]
\input{fitted-flowcurves.tex}
\end{scope}

\begin{scope}[xshift=0.7\textwidth,]
\input{fit-parameters.tex}
\end{scope}

\end{tikzpicture} 
\caption{Constitutive relation of an air-fluidized granular fluid. (A)
  Dimensionless shear stress $\sigma/nT_0$ as a function of
  dimensionless true shear rate $\dot\gamma/\omega_0$ on
  double-logarithmic axes. With this normalization, all temperature
  dependence is absorbed in the ideal gas pressure $nT_0$ and the
  collision frequency $\omega_0$ (Fig.~\ref{fig:shearbanding}C) of the
  unsheared fluid. Symbols are the measurements (labeled by the
  fluidization velocity $u$; for the corresponding packing fractions
  see Tab.~\ref{tab:values}) and lines are the GITT fits
  (Fig.~\ref{fig:shearbanding}). In this representation, there is only
  one fit parameter per curve, the distance to the granular glass
  transition $\Delta\varphi_\mathrm{th}$
  (Fig.~\ref{fig:shearbanding}D). (B) All the measurements (symbols)
  located in the rheological state diagram spanned by the
  dimensionless shear rate and $\Delta\varphi_\mathrm{th}$. For the
  color code see Fig.~\ref{fig:experiment}. (C) Compressibility factor
  $Z$ as a function of packing fraction calculated for the inner and
  outer region of the shear band showing that the fluidized bed
  features a unique equation of state. Same symbols as in A.}
  \label{fig:gitt}
\end{figure*}

\subsection*{Shear Banding}

The Taylor-Couette geometry enforces a radial variation of the stress
$\sigma(r)\sim r^{-2}$ (see \gls{si}). For a measured stress only
slightly above the stress plateau in the flow curve, this implies a
large variation of the shear rate $\dot\gamma(r)$ in a small radial
interval. Given the finite size particles, this is compatible with the
interpretation of the non-monotonicity of $\sigma(\dot\gamma_e)$ as a
manifestation of shear banding. More precisely, we assume that the
sheared fluidized bed may radially split into an \emph{inner} ($r < r^*$) and
an \emph{outer} region ($r > r^*$), which are in different states and with a
discontinuous jump of the shear rate at the interface between the
regions at $r = r^*$.

To account for shear banding in \gls{gitt}, we allow for different
sets of parameters in the outer and inner region of the
shear cell, ($nT_0^{\mathrm{o}}$, $\omega_0^{\mathrm{o}}$,
$\Delta\varphi^{\mathrm{o}}$), and ($nT_0^{\mathrm{i}}$,
$\omega_0^{\mathrm{i}}$, $\Delta\varphi^{\mathrm{i}}$),
respectively. Moreover, we assume no-slip boundary conditions, i.e.,
in terms of the local turning rate $U(r=R_\mathrm{i}) \equiv U$, and
$U(r=R_\mathrm{o})\equiv 0$. As we know the
strain constants outside the shear banding regime, and local shear
rates, $\dot\gamma(r)$, necessarily 
decrease from the inside to the outside,
we fit the outer set of parameters to the low shear rate
section and the inner set of parameters to the high shear rate section
of the experimental flow curve, without regard to non-monotonous
section at intermediate shear rates. However, the two branches of the
constitutive equation needs to be unique for asymptotically large and
small shear rates (see Fig.~\ref{fig:shearbanding}A for an example and
Fig.~S2 in \gls{si} for all fits). In the shear banding regime, we
assume a coexistence of materials described by the two sets of
parameters with an interface at a radial position
$R_{\mathrm{i}} < r^* < R_{\mathrm{o}}$ (see
Fig.~\ref{fig:shearbanding}B and Fig.~S6), where $r^*$ is determined
by imposing continuity of stress and turning rate at the interface
(see Methods for more details).

This procedure equips us with a lot of physical information, namely
estimates for the granular temperature $T_0^{\mathrm{i,o}}$ and the
collision frequency $\omega_0^{\mathrm{i,o}}$
(Fig.~\ref{fig:shearbanding}C), as well as theoretical estimates of
the distance to the glass transition $\Delta\varphi^{\mathrm{i,o}}$
(Fig.~\ref{fig:shearbanding}D).  We will come back to those in the
Discussion, but for now, let us focus on the shear rate profiles
(Fig.~\ref{fig:shearbanding}B). In particular, we can read off the
true shear rate at the inner cylinder
$\dot\gamma = \dot\gamma(r = R_{\mathrm{i}})$. Plotting the latter
against the effective shear rate $\dot\gamma_{\mathrm{e}}$, as in
Fig.~\ref{fig:shearbanding}E, we observe that the fitting procedure
conforms to the analytical strain constants at low and high shear
rates, and that in the shear-banding regime, the relation is indeed
highly non-linear.  Although we only measure the stress at the inner
cylinder, the stress at the outer cylinder, $\sigma/\delta^2$, is
fixed by the force balance (see \gls{si}). With the corresponding
shear rate $\dot\gamma = \dot\gamma(r = R_{\mathrm{o}})$, this allows
us to extract a second flow curve from every measurement.

Given that we have obtained the relevant stress and rate scales,
$nT_0$ and $\omega_0$, as well as the true shear rate $\dot\gamma$, we
can replot the experimental flow curves in the proper dimensionless
form, $\sigma(\dot\gamma/\omega_0)/nT_0$, using the true shear rate
$\dot\gamma$ at both the inner and outer cylinder
(Fig.~\ref{fig:gitt}A). Comparison to the \gls{gitt} constitutive
relation shows excellent agreement over many orders of magnitude in
shear rate and stress. Note that in this representation,
$\Delta\varphi_\text{th}$ is the single fit parameter per flow curve.

\section*{Discussion}
\label{sec:discussion}

The constitutive relation, $\sigma(\dot\gamma)$, for an air-fluidized
granular bed, featured in Fig.~\ref{fig:gitt}A,
confirms that the manifold rheology of fluidized granular
media, in fact, displays different rheological regimes: (i) For small
Weissenberg number, $\Wi\ll1$, and negligible shear heating, the
fluidized bed behaves as a Newtonian fluid with a viscosity $\eta_\text{N}$
(Fig.~\ref{fig:measurements} and Tab.~\ref{tab:values}) that diverges
towards the granular glass transition at an effective packing fraction
$\varphi_g = \phig$. (ii) For higher shear rates, $\Wi\gtrsim 1$,
shear thinning and, eventually, at the granular glass transition
$\varphi_\mathrm{g} = \phig$, a yield
stress $\sigma_0\approx\SI{5}{\pascal}$ is observed as the shear rate
exceeds the structural relaxation rate of the granular fluid. (iii)
Once shear heating dominates, the rheology crosses over into Bagnold
scaling, $\sigma = B{\dot\gamma}^2$, characterized by the Bagnold
coefficient $B$ (Fig.~\ref{fig:measurements} and
Tab.~\ref{tab:values}).

It is remarkable that the shape of the constitutive relations are
solely controlled by $\Delta\varphi_\text{th}$, the distance to the granular
glass transition. Note that, in order to cover all rheological regimes
(Fig.~\ref{fig:gitt}B), measurements had to be performed consistently
over many orders of magnitude in shear rate as well as shear stress,
and over long times.

The transition from Newtonian to shear shinning at $\Wi\sim1$ is
generic for dense suspensions \cite{Stickel2005,Fuchs2002} as all
suspensions (and, in fact, all fluids) feature a structural relaxation
time $\tau$. The location of the glass transition, $\varphi_\text{g}$,
however, depends on the specifics of the particle interactions. 
While dense suspensions generally feature a shear thickening regime at
high shear rates \cite{Stickel2005,Brown2014}, the Bagnold regime of
the fluidized bed is probably of different origin. Rather than
frictional or hydrodynamic interactions increasing the suspension's
viscosity as in the case of over-damped suspensions, the inertial
granular fluidized bed features an increasing viscosity in the Bagnold
regime due to the increasing granular temperature. The origin of this
phenomenon is the decoupling of the granular temperature of the
particles from the thermodynamic temperature of the suspending fluid
that acts as a thermostat in more conventional suspension
\cite{Lemaitre2009,Kranz2020}.
As we are studying air-fluidized glass, beads, the
viscosity $\eta$ is always orders of magnitude larger than the air's
viscosity $\eta_f$, such that we observe no viscous-inertial transition
\cite{Madraki2020,Tapia2022}.

At first sight, our measured granular glass transition volume fraction
$\varphi_\mathrm{g} = \phig$ is surprisingly low. For colloids in thermal
equilibrium, $\varphi_\mathrm{g}\simeq0.57$--$0.58$ \cite{vanMegen1995}.
\Gls{gmct} predicts that $\varphi_\mathrm{g}$ should
increase further the more dissipative particles are \cite{Kranz2010}.
In particular, $\varphi_\mathrm{g}^\mathrm{MCT}(\varepsilon=0.8) = 0.528$
compared to $\varphi_\mathrm{g}^\mathrm{MCT}(\varepsilon=1) = 0.517$ with
a well-known offset relative to experimental values. 
Note that $\varphi_g$ is
defined by the apparent divergence of the Newtonian viscosity at
infinitesimal shear rate, $\Wi\ll1$. Here, the fluid's structure has
sufficient time to adapt to the shear flow, such that this is a
phenomenon distinct from the proposed shear jamming behavior found at
much higher packing fractions and for $\Wi\gg1$
\cite{BiZhang2011}. This adds further evidence, that a particulate
fluid features a glass \emph{and} a jamming transition
\cite{Charbonneau2017,Kranz2010} rather than those two being different
manifestations of a single transition \cite{Liu1998}.

Another manifestation of the low Stokes number, lubrication dominated
particle interactions is the fact that the coefficient of restitution
becomes strongly dependent on the Stokes number,
$\varepsilon = \varepsilon(\St)$, \cite{Gondret2002,Yang2008} and is
no longer a material property of the glass beads. In particular,
$\varepsilon = 0$ for $\St\lesssim 10$. While at speeds on the order
of \si{\metre\per\second}, glass spheres have
$\varepsilon=\varepsilon_{\infty}\gtrsim0.87$ \cite{Lorenz1997}, this
implies that for the collision speeds relevant in the fluidized bed,
the coefficient of restitution becomes strongly speed dependent,
$\varepsilon = \varepsilon(v)$, with a typical value,
$\varepsilon(v_0)$ much smaller than the high speed value
$\varepsilon_{\infty}$. The data analysis in terms of \gls{gitt}
requires to specify an effective coefficient of restitution for the
fluidized bed as a whole. The procedure works for any value of
$\epsilon$ (see \gls{si}) with the result that the distance to the
glass transition, $\Delta\varphi_\mathrm{th}$, and the stress scale,
$nT_0$, only weakly depend on $\varepsilon$, whereas the rate scale,
$\omega_0$, displays a marked dependence on $\varepsilon$.

Finally lubrication in combination with packing fractions far below
jamming explains why we do not have to take into account
inter-particle friction \cite{DeGiuli2016}. While it is likely that
residual frictional contacts renormalize the inferred system
parameters to some extent, no new phenomenology appears that isn't
captured by \gls{gitt}, which assumes frictionless particles.

In the fully fluidized state, the
granular temperatures increase with fluidization speed as
expected. Typical particle speeds, inferred from the granular
temperature $T_0$ are of the order but smaller than the fluidization
velocity, $v_0\lesssim u$, as expected \cite{Biggs2008}. The opposite
trend in the partially fluidized state hints at additional
contributions to the stress scale $nT_0$, likely borne by lasting
particle contacts.
Interestingly, the granular temperature in the inner region is
significantly lower than in the outer region
(Fig.~\ref{fig:shearbanding}C). This implies that strong shear somehow
lowers the efficiency of fluidization, a phenomenon that certainly
warrants further investigation.

Neither the stress scale, $nT_0$, nor the rate scale, $\omega_0$, can
be easily measured independently in a fluidized bed
\cite{Cody1996,Biggs2008, AguilarCorona2011}, and doing so is beyond
the scope of the present study.

Predicting the expected collision frequency $\omega_0$ suffers from
the notorious lack of a good theoretical equation of state,
$P(\varphi) = nT_0Z(\varphi)$, giving the density dependence of the
particle pressure $P$ for the relevant packing fractions of order
$0.5$. 
For typical particle speeds,
$v_0\lesssim\SI{2}{\centi\metre\per\second}$, and assuming the
Carnahan-Starling EOS, $Z_{\mathrm{CS}}(\varphi)$
\cite{Carnahan1969}, we estimate $\omega_0\lesssim\SI{10}{\kilo\hertz}$
in line with the collision frequency extracted from the measurements
(Fig.~\ref{fig:shearbanding}C). Given the significant
$\varepsilon$-dependence of $\omega_0$, this led us to settle on an
effective value for the coefficient of restitution $\varepsilon=0.8$.
For $\varepsilon=0.8$, the parameters 
$nT_0$ and $\omega_0$
assume reasonable values, as shown in
Figs.~\ref{fig:shearbanding} \& \ref{fig:gitt}. Note that both
parameters reflect the transition from
the partially, to the fully fluidized state which occurs at 
$u\approx\SI{0.01}{\meter\per\s}$ (Fig.~\ref{fig:shearbanding}C).

Although the parameters of the
constitutive relation in the inner and outer region of the shear band
differ (Fig.~\ref{fig:shearbanding}C), the equation of state of the
unsheared fluidized bed, $Z(\varphi)$, must be unique throughout the
sample. The compressibility factor, $Z(\varphi)$, can be estimated
from the stress and rate scales (see Methods) for the inner and outer
region. Fig.~\ref{fig:gitt}C shows reasonable data collapse,
confirming a unique equation of state.

Note that for the present study we have explicitly taken a macroscopic
point of view and have not used spatially resolved measurements. In
particular, we have based our shear banding analysis on two
simplifying assumptions: (i) that the inner and the outer region are
both always in a sheared fluidized state, and (ii), that the regions
are joined at a sharp interface. In the experiment, the interface will
necessarily have a certain width on the particle scale and shear
banding could also manifest itself in localized shear, leaving part of
the sample unsheared. It would be interesting to measure particle
motion in the bulk of the sample with sufficient temporal and spatial
resolution to directly observe the flow profiles and compare them to
the those determined here
based on \gls{gitt}. 
While challenging, we expect this to be in reach of recent
advances in the field of \textit{in-situ} visualization techniques~\cite{Baker2018, Rosato2020, Stannarius2017}.
More demanding would be to achieve the resolution necessary to
independently determine particle speeds or collision frequencies;
measurements that would be very welcome, as well, in light of the role
that those quantities play for the constitutive relation of a
fluidized bed.

Apart from the radial separation into an inner and outer region,
respectively, we have treated the sample as a spatially homogeneous
material by time-averaging over the dynamics of the bubbles. We
imagine that, eventually, our constitutive relation can be generalized
from the \emph{global} scale embraced here to a
\emph{local} -- spatially and temporally resolved -- continuum
description of the dynamic, bubbling fluidized bed \cite{Guo2021}
including start-up flows \cite{Gnoli2016}. Of particular conceptual
interest would be a thorough analysis of the flow instabilities as has
been done for molecular fluids~\cite{Taylor1923,Stuart1986} but is, to
date, mostly unexplored for granular fluids~\cite{Conway2004,Krishnaraj2016}.

Let us stress that our proposed constitutive relation
does not contradict the
$\mu(\mathcal I)$-rheology \cite{Midi2004} or its generalizations to
suspensions \cite{Boyer2011} but rather includes it
\cite{Lemaitre2009,Coquand2020} and extends it beyond the Bagnold
regime, where shear heating is no longer dominant and the fluidization
flow becomes important.

In summary, we have validated and calibrated the \gls{gitt}
constitutive relation as a faithful model of the manifold rheology of
air fluidized glass beads. On an explanatory level, this implies that
fluidized beds will display Newtonian rheology if the shear rate is
small compared to the intrinsic relaxation rate of the particle bed,
$\Wi\ll1$, and shear heating can be neglected; they will display shear
thinning for $\Wi\gg1$, as long as shear heating remains negligible;
and finally, they will cross over to Bagnold scaling once shear heating
dominates over fluidization. We hope that our work helps in
understanding industrial and geophysical granular flows and can
eventually lead to faithful continuum simulations \cite{Wang2020} of
large-scale granular applications.

\matmethods{
\subsection*{Rheometry}
All measurements presented were obtained using an
Anton Paar MCR 302 rheometer with a powder cell and profiled inner cylinder (Fig.~\ref{fig:rheometerMethod}A),
as described in Ref.~\cite{Salehi2018}.
The geometry of the setup is an open-surface Taylor-Couette;
important dimensions are given in Fig.~\ref{fig:rheometerMethod}B.

\begin{figure}[h!]
\centering
\input{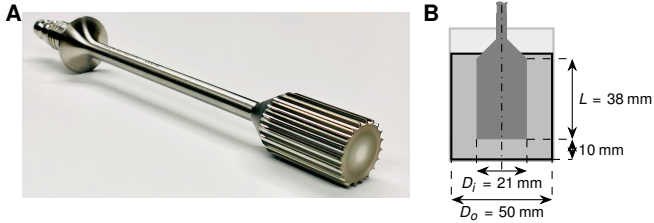}
\caption{\label{fig:rheometerMethod}
Rheometry setup. (A)~Profiled inner cylinder. (B)~Important dimensions for the Taylor-Couette shear cell. 
}
\end{figure}

The rheometer records the torque $\mathcal M$ as a function of the
inner cylinder's rotation rate $U$. The torque can easily be converted
to a stress at the inner cylinder, $\sigma = \mathcal M/\pi
LR_i^2$. For purely Newtonian fluids in a Taylor-Couette geometry, the
effective shear rate $\dot\gamma \equiv \dot\gamma_e = K_N2\pi U$ is
linearly related to the turning rate $U$ by the strain constant
\begin{equation}
  \label{eq:4}
  K_N = \frac{2\delta^2}{\delta^2 - 1}.
\end{equation}
For all other rheologies, $\dot\gamma_\mathrm{e}$, designates an
effective shear rate, different from the true shear rate $\dot\gamma$.

\subsection*{Measurement Procedure}
The flow curves presented in Fig.~\ref{fig:experiment}G are
measured using the following experimental procedure: For a given
fluidization velocity $u$, i.e.~packing fraction $\varphi$, a desired apparent shear
rate $\dot\gamma_\text{e}$ is imposed. Sufficient preshear is applied to
reach a stationary state, as evidenced by reduced stress fluctuating
around a constant mean value. Note that this takes a long
time (up to strain $\gamma \approx 200$), as expected for a yield stress fluid \cite{Moller2009}. 
The mean value $\sigma$ is obtained by averaging the stress over an additional
strain window of twice the strain necessary to reach steady state.

The mean $\sigma$ hence constitutes one
data point in Fig.~\ref{fig:experiment}G, the error bar representing
the standard deviation $\delta\sigma$ over the stationary state strain
window. Between individual data points, the fluidization velocity $u$
is increased to \SI{0.05}{\m\per\s} to erase
any memory of the previous shear states. To check consistency, flow
curves are measured successively in an upward and downward
sweep in shear rate (respectively filled and empty marks in
Figs.~\ref{fig:experiment}G, \ref{fig:measurements}A and \ref{fig:shearbanding}D).
For each flow curve, each point $\sigma(u,\,\dot{\gamma}_\text{e})$
is measured consecutively to ensure continuity in all flow curves,
resulting in a measuring time per flow curve of around
\SI{200}{\hour}.

\subsection*{Particles}
\label{sec:particles}

All measurements use a sample of mass $M=\SI{125}{g}$ of
soda-lime glass spherical beads from Mo-Sci Corp,
of diameter
\SIrange{63}{75}{\micro\metre}
(mean diameter $d=\SI{69}{\micro\metre}$).
The bulk material density is $\rho_\mathrm{b} =
\SI{2.5}{\g\per\cubic\centi\metre}$. 
The same glass beads were characterized in prior publications,
see Refs.~\cite{LaMarche2016, Mishra2020}.

\subsection*{Packing Fraction}
The packing fraction is measured globally: it is calculated as
$\varphi = M / \rho_\text{b} Ah$, where the sample volume $Ah$ is
obtained from the cross section of the shear cell, $A$, and the mean
height $h$ of the fluidized bed as determined by image analysis. The
uncertainty in $\varphi$ is dominated by systematic uncertainty in $h$
which we estimate at $\pm2.7\%$. For large air flow velocities $u$,
fluctuation in $h$ due to bubbling instabilities also becomes
significant. The fluctuation magnitude is represented as error bars in
Fig.~\ref{fig:experiment}B (standard deviation in $\varphi$ for five
independent repetition).

For the data presented in
Fig.~\ref{fig:experiment}A and Fig.~\ref{fig:experiment}B, air flow
velocity $u$ and packing fraction $\varphi$ are measured without the
inner cylinder. The relation $\varphi(u)$ is obtained by fitting this
data with the expression:
\begin{equation}
  \varphi(u) = \varphi_{0} + \frac{ \varphi_{\infty} - \varphi_{0} }{
    1 + (\nicefrac{u_0}{(u - u_f)})^\exposant}
\end{equation}
with $ \varphi_{\infty} = \phiRLP$, $\varphi_{0} = \phiINF$,
$u_0 = \SI{\uzero}{\m\per\s}$ and
$u_f = \SI{\ufp}{\m\per\s}$.
The fit is represented in Fig.~\ref{fig:experiment}B by a dashed gray line.

\subsection*{Fluidization Curves}
\label{sec:fluidization-curves}

The excess pressure $\Delta p$ relative to ambient air pressure is measured at
the base of the fluidized bed by a pressure transducer (mass flow controller from B\"{u}rkert).

\subsection*{Granular Integration Through Transients}
\label{sec:gran-integr-thro}

To put it succinctly,
\gls{gitt}~\cite{Kranz2018,Kranz2020} expresses the
dimensionless shear stress $\hat\sigma = \sigma/nT$ of a dense fluid
of dissipative smooth hard spheres as a generalized Green-Kubo
integral
\begin{equation}
  \label{eq:3}
  \hat\sigma(\Pe) = \Pe\sum_{\vec q}\int_0^{\infty}dt
  \mathcal V_{\vec q\vec q(-t)}\Phi^2_{\vec q(-t)}(t),
\end{equation}
where $\Pe = \dot\gamma/\omega_c$ is a P\'eclet number. The physical
intuition behind this equation is, that of all stress relaxation
modes, the slowest one in dense fluids will be the density
fluctuations encoded in the respective density correlator
$\Phi_{\vec q}(t)$ \cite{Fuchs2002}. The stress-density coupling
constant $V_{\vec q\vec q(-t)}$ is known explicitly
\cite{Kranz2020}. In addition to the explicit shear rate dependence of
the above relation, the shear rate also affects the advection of the
wave vectors, $\vec q(-t)$, and the density correlator
$\Phi_{\vec q}(t)$ allowing for non-Newtonian rheology
\cite{Fuchs2002,Kranz2020}. For granular fluids, as opposed to
thermostatted colloidal suspensions, shear heating,
$\sigma\dot\gamma$, elevates the granular temperature $T > T_0$
compared to the granular temperature $T_0$ of the unsheared granular
fluid. The relation $T(\Pe)/T_0$ is known \cite{Kranz2020}.

The \gls{gitt} consititutive equation,
$\hat\sigma(\Pe) = \hat\sigma(\Pe|\varphi, \epsilon)$, uses the
packing fraction $\varphi$ and the coefficient of restitution
$\epsilon$ to uniquely characterize the granular fluid. Note, however,
that the dominant effect of $\epsilon$ is in changing the glass
transition density $\varphi_\mathrm{g} = \varphi_\mathrm{g}^\mathrm{MCT}(\epsilon)$
\cite{Kranz2010} and only the distance to the glass transition
$\Delta\varphi_\mathrm{th} = \varphi_\mathrm{g}^\mathrm{MCT}(\epsilon) - \varphi$ matters for
$\hat\sigma$. Therefor we regard the \gls{gitt} constitutive relation,
\begin{equation}
  \label{eq:5}
  \sigma(\Pe\sqrt{T(\Pe)/T_0}\mid\Delta\varphi_\mathrm{th})/nT_0
  = \hat\sigma(\Pe\mid\Delta\varphi_\mathrm{th})T(\Pe)/T_0,
\end{equation}
as characterized by the stress and rate scale, $nT_0$, and $\omega_0 =
\omega_c(T_0)$, respectively, as well as one more parameter, namely
$\Delta\varphi_\mathrm{th}$.

\subsection*{Fitting the Shear Band}

We assume no-slip boundary conditions at the inner and outer cylinder
and also at the shear band. Assuming two different \gls{gitt}
constitutive relations, $\sigma^i(\dot\gamma),\sigma^o(\dot\gamma)$,
at the inner and outer part of the shear cell, respectively, that meet
at radial coordinate $r^*$, Eq.~\textbf{S2} (see \gls{si}) generalizes to a set of
two equations
\begin{equation}
  \label{eq:2}
  \begin{gathered}
    \int_{\sigma/\delta^2}^{\sigma^*}\frac{{\dot\gamma}^{\mathrm{o}}(s)ds}{2s} = 2\pi U^*\\
    \int_{\sigma^*}^{\sigma}\frac{{\dot\gamma}^{\mathrm{i}}(s)ds}{2s}
    = 2\pi(U - U^*)
  \end{gathered}
\end{equation}
for the two unknowns, $\sigma^*$ and $U^*$, the shear stress and
turning rate at the shear band, respectively. Here
${\dot\gamma}^{\mathrm{i,o}}(\sigma)$ are the inverse of the
constitutive relations $\sigma^{\mathrm{i,o}}(\dot\gamma)$. Note that
$\sigma^*$ is uniquely related to the location $r^*$ by the force
balance (see \gls{si}).

We also need to determine the parameters of the \gls{gitt}
constitutive relations. As we do not \textit{a priori} know the
relation $\dot\gamma(U)$ in the non-monotonic middle part of the flow
curves, we skip this part in the fitting procedure and assume the
ideal strain constants only at the low and high shear rate ends of the
flow curves. We fix the coefficient of restitution $\epsilon$ for all
fits and then adjust the free parameters, $nT_0^{\mathrm{i,o}}$,
$\omega_0^{\mathrm{i,o}}$, and $\Delta\varphi^{\mathrm{i,o}}$ until we
obtain a visually satisfying fit (see Fig.~\ref{fig:shearbanding} and
Figs.~S2--5). Note that we require the collision frequencies to scale
consistently with the temperatures,
$\omega_0^{\mathrm{i}}/\omega_0^{\mathrm{o}} =
\sqrt{T_0^{\mathrm{i}}/T_0^{\mathrm{o}}}$. This procedure fixes the
constitutive relations $\sigma^{\mathrm{i,o}}(\dot\gamma)$, valid in
the Bagnold and Newtonian regime, respectively. Next, we solve
Eqs.~\textbf{\ref{eq:2}} for the location of the shear band for those data
points that deviate from the fit by either $\sigma^i(\dot\gamma)$ or
$\sigma^o(\dot\gamma)$ alone.

By this procedure, we can associate every measurement
$(U, \sigma(U))$, with a shear rate
$\dot\gamma^i\equiv\dot\gamma^i(\sigma(U))$ at the inner cylinder and
a shear rate $\dot\gamma^o\equiv\dot\gamma^o(\sigma(U)/\delta^2)$ at
the outer cylinder. In the main part, we fix $\epsilon=0.8$ and fit to
the up-sweep measurements. For other choices of $\epsilon$ and for the
down-sweep measurements, see the \gls{si}.

\subsection*{Determining the compressibility factor}
\label{sec:determ-compr-fact}
The collision frequency of hard spheres can be written as a function
of $Z$,
$\omega_0 = \omega_0(Z, T_0) = 6\sqrt{T_0/\pi m}[Z(\varphi)-1]/d$
\cite{Poeschel2001}, which can be inverted to obtain
$Z^{i,o}(\omega_0^ {i,o}, T_0^{i, o})$.

\subsection*{Data Availability}
All data used in the paper is publicly available as \gls{si}.
}

\showmatmethods{} 

\acknow{We are indebted to Dennis Schütz for designing the powder cell
  and for sharing unpublished measurements that were the starting
  point of this study. We thank Matthias Fuchs, Annette Zippelius, and
  Olivier Coquand for companionship in the long development of
  \gls{gitt} and many more interesting discussions. We thank Thomas
  Voigtmann for sharing rheological insight and steady
  encouragement.
W.\,T.\,K.~acknowledges funding by the DFG through grant
  number KR4867/2 and the German Aerospace Centre through the project
  SciML. 
O.\,D'A.~acknowledges funding by 
DLR/DAAD Research Fellowship 91647576
\& ESA NPI-4000122340.
}

\showacknow{}

\bibliography{bibliography.bib}

\begin{thebibliography}{10}

\bibitem{Geldart1973}
D Geldart, Types of gas fluidization.
\newblock {\em\protect\JournalTitle{Powder Technology}} \textbf{7}, 285--292 (1973).

\bibitem{FluidizationEngineeringBook1991}
D Kunii, O Levenspiel, eds., {\em Fluidization Engineering}.
\newblock (Elsevier Science), 2 edition, (1991).

\bibitem{Zenz1997}
FA Zenz, {\em Fluidization Phenomena and Fluidized Bed Technology}, eds.{} ME Fayed, L Otten.
\newblock (Springer US, Boston, MA), pp. 487--531 (1997).

\bibitem{Horio2017}
M Horio, Fluidization in natural phenomena in {\em Reference Module in Chemistry, Molecular Sciences and Chemical Engineering}.
\newblock (2017).

\bibitem{Goldhirsch2008}
I Goldhirsch, Introduction to granular temperature.
\newblock {\em\protect\JournalTitle{Powder Technology}} \textbf{182}, 130--136 (2008).

\bibitem{Bakhtiyarov1999}
S Bakhtiyarov, R Overfelt, Recent advances in the rheology of fluidized materials in {\em Advances in the Flow and Rheology of Non-Newtonian Fluids}, Rheology Series, eds.{} D Siginer, D {De Kee}, R Chhabra.
\newblock Vol.{}~8, pp. 1399--1433 (1999).

\bibitem{Roche2004}
O Roche, M Gilbertson, J Phillips, R Sparks, Experimental study of gas-fluidized granular flows with implications for pyroclastic flow emplacement.
\newblock {\em\protect\JournalTitle{{Journal of Geophysical Research}}} \textbf{109} (2004).

\bibitem{Kottlan2018}
A Kottlan, D Sch{\"u}tz, S Radl, Rheological investigations on free-flowing and cohesive powders in different states of aeration, using a ball measuring system.
\newblock {\em\protect\JournalTitle{Powder Tech.}} \textbf{338}, 783--794 (2018).

\bibitem{Breard2022}
ECP Breard, et~al., Investigating the rheology of fluidized and non-fluidized gas-particle beds: implications for the dynamics of geophysical flows and substrate entrainment.
\newblock {\em\protect\JournalTitle{Granular Matter}} \textbf{24}, 34 (2022).

\bibitem{Dijksman2011}
JA Dijksman, GH Wortel, LTH van Dellen, O Dauchot, M van Hecke, Jamming, yielding, and rheology of weakly vibrated granular media.
\newblock {\em\protect\JournalTitle{Phys. Rev. Lett.}} \textbf{107}, 108303 (2011).

\bibitem{Zhang2018}
Z Zhang, Y Cui, DH Chan, KA Taslagyan, {DEM} simulation of shear vibrational fluidization of granular material.
\newblock {\em\protect\JournalTitle{Granular Matter}} \textbf{20}, 71 (2018).

\bibitem{Gnoli2016}
A Gnoli, A Lasanta, A Sarracino, A Puglisi, Unified rheology of vibro-fluidized dry granular media: From slow dense flows to fast gas-like regimes.
\newblock {\em\protect\JournalTitle{Scientific Reports}} \textbf{6}, 38604 (2016).

\bibitem{Melosh1996}
HJ Melosh, Dynamical weakening of faults by acoustic fluidization.
\newblock {\em\protect\JournalTitle{Nature}} \textbf{379}, 601--606 (1996).

\bibitem{Conrad2013}
JW Conrad, J Melosh, The rheology of acoustically fluidized sand in {\em AGU Fall Meeting Abstracts}.
\newblock Vol.{} 2013, pp. P41F--1986 (2013).

\bibitem{Wilson1984}
C Wilson, The role of fluidization in the emplacement of pyroclastic flows, 2: Experimental results and their interpretation.
\newblock {\em\protect\JournalTitle{Journal of Volcanology and Geothermal Research}} \textbf{20}, 55--84 (1984).

\bibitem{Kelfoun2009}
K Kelfoun, P Samaniego, P Palacios, D Barba, Testing the suitability of frictional behaviour for pyroclastic flow simulation by comparison with a well-constrained eruption at tungurahua volcano (ecuador).
\newblock {\em\protect\JournalTitle{Bull. Volcanol.}} \textbf{71}, 1057--1075 (2009).

\bibitem{Eames2005}
I Eames, M Gilbertson, Mixing and drift in gas-fluidised beds.
\newblock {\em\protect\JournalTitle{Powder Technology}} \textbf{154}, 185--193 (2005).

\bibitem{Werther2007}
J Werther, {\em Fluidized-Bed Reactors}.
\newblock (John Wiley \& Sons, Ltd), (2007).

\bibitem{Menendez2019}
M Men{\'e}ndez, J Herguido, A B{\'e}rard, GS Patience, Experimental methods in chemical engineering: Reactors---fluidized beds.
\newblock {\em\protect\JournalTitle{The Canadian Journal of Chemical Engineering}} \textbf{97}, 2383--2394 (2019).

\bibitem{Dhurandhar2018}
R Dhurandhar, JP Sarkar, B Das, The recent progress in momentum, heat and mass transfer studies on pneumatic conveying: a review.
\newblock {\em\protect\JournalTitle{Heat and Mass Transfer}} \textbf{54}, 2617--2634 (2018).

\bibitem{Matheson1949}
GL Matheson, WA Herbst, PH Holt, Characteristics of fluid-solid systems.
\newblock {\em\protect\JournalTitle{Ind. Eng. Chem.}} \textbf{41}, 1098--1104 (1949).

\bibitem{Hagyard1966}
T Hagyard, AM Sacerdote, Viscosity of suspensions of gas-fluidized spheres.
\newblock {\em\protect\JournalTitle{Ind. Eng. Chem. Fundamentals}} \textbf{5}, 500--508 (1966).

\bibitem{Maeno1980}
N Maeno, K Nishimura, Y Kaneda, Viscosity and heat transfer in fluidized snow.
\newblock {\em\protect\JournalTitle{J. Glaciol.}} \textbf{26}, 263--274 (1980).

\bibitem{Koos2012}
E Koos, E Linares-Guerrero, ML Hunt, CE Brennen, Rheological measurements of large particles in high shear rate flows.
\newblock {\em\protect\JournalTitle{Phys. Fluids}} \textbf{24}, 013302 (2012).

\bibitem{Bouwhuis1961}
P van~den Leeden, GJ Bouwhuis, Tentative rules for shearing stresses in particulate fluidized beds.
\newblock {\em\protect\JournalTitle{Appl. Sci. Res.}} \textbf{10}, 78--80 (1961).

\bibitem{Tardos1998}
GI Tardos, MI Khan, DG Schaeffer, Forces on a slowly rotating, rough cylinder in a couette device containing a dry, frictional powder.
\newblock {\em\protect\JournalTitle{Phys. Fluids}} \textbf{10}, 335--341 (1998).

\bibitem{Hanes1985}
DM Hanes, DL Inman, Observations of rapidly flowing granular-fluid materials.
\newblock {\em\protect\JournalTitle{J. Fluid Mech.}} \textbf{150}, 357--380 (1985).

\bibitem{Schugerl1961}
K Sch{\"u}gerl, M Merz, F Fetting, Rheologische eigenschaften von gasdurchstr{\"o}mten fliessbettsystemen.
\newblock {\em\protect\JournalTitle{Chem. Engng. Sci.}} \textbf{15}, 1--38 (1961).

\bibitem{Hobbel1985}
EF Hobbel, B Scarlett, Measurement of the flow behaviour of aerated and fluidised powders using a rotating viscometer.
\newblock {\em\protect\JournalTitle{Particle \& Particle Systems Characterization}} \textbf{2}, 154--159 (1985).

\bibitem{Gottschalk1986}
J Gottschalk, Rheological study of loosened bulk granular materials.
\newblock {\em\protect\JournalTitle{Particle \& Particle Systems Characterization}} \textbf{3}, 168--173 (1986).

\bibitem{Mishra2020}
I Mishra, P Liu, A Shetty, CM Hrenya, On the use of a powder rheometer to probe defluidization of cohesive particles.
\newblock {\em\protect\JournalTitle{Chemical Engineering Science}} \textbf{214}, 115422 (2020).

\bibitem{Young2021}
AB Young, A Shetty, ML Hunt, Flow transitions and effective properties in an annular couette rheometer for gas fluidized beds and liquid-solid suspensions.
\newblock {\em\protect\JournalTitle{submitted to J. Fluid Mech.}} (2021).

\bibitem{Bagnold1954}
RA {Bagnold}, Experiments on a gravity-free dispersion of large solid spheres in a newtonian fluid under shear.
\newblock {\em\protect\JournalTitle{Proceedings of the Royal Society of London Series A}} \textbf{225}, 49--63 (1954).

\bibitem{Lois2005}
G Lois, A Lema{\^\i}tre, JM Carlson, Numerical tests of constitutive laws for dense granular flows.
\newblock {\em\protect\JournalTitle{Phys. Rev. E}} \textbf{72}, 051303 (2005).

\bibitem{Midi2004}
{{GDR} {MiDi}}, On dense granular flows.
\newblock {\em\protect\JournalTitle{European Physical Journal E}} \textbf{14}, 341--365 (2004).

\bibitem{Francia2021}
V Francia, LAA Yahia, R Ocone, A Ozel, From quasi-static to intermediate regimes in shear cell devices: Theory and characterisation.
\newblock {\em\protect\JournalTitle{KONA Powder Particle J.}} p. 2021018 (2021).

\bibitem{Kranz2018}
WT Kranz, F Frahsa, A Zippelius, M Fuchs, M Sperl, Rheology of inelastic hard spheres at finite density and shear rate.
\newblock {\em\protect\JournalTitle{Physical Review Letters}} \textbf{121}, 148002 (2018).

\bibitem{Kranz2020}
WT Kranz, F Frahsa, A Zippelius, M Fuchs, M Sperl, Integration through transients for inelastic hard sphere fluids.
\newblock {\em\protect\JournalTitle{Physical Review Fluids}} \textbf{5}, 024305 (2020).

\bibitem{Lemaitre2009}
A Lema{\^\i}tre, JN Roux, F Chevoir, What do dry granular flows tell us about dense non-brownian suspension rheology?
\newblock {\em\protect\JournalTitle{Rheol. Acta}} \textbf{48}, 925--942 (2009).

\bibitem{Guazzelli2018}
{\'E} Guazzelli, O Pouliquen, Rheology of dense granular suspensions.
\newblock {\em\protect\JournalTitle{J. Fluid Mech.}} \textbf{852} (2018).

\bibitem{Wang2020}
J Wang, Continuum theory for dense gas-solid flow: A state-of-the-art review.
\newblock {\em\protect\JournalTitle{Chem. Engng. Sci.}} \textbf{215}, 115428 (2020).

\bibitem{LaMarche2016}
CQ LaMarche, P Liu, KM Kellogg, CM Hrenya, Fluidized-bed measurements of carefully-characterized, mildly-cohesive (group a) particles.
\newblock {\em\protect\JournalTitle{Chem. Engng. J.}} \textbf{310}, 259--271 (2016).

\bibitem{Grace2008}
JR Grace, B Leckner, J Zhu, Y Cheng, {\em Fluidised Beds}, ed.{} CT Crow.
\newblock (CRC Press), pp. 5--71 (2008).

\bibitem{Valverde1998}
JM Valverde, A Ramos, A Castellanos, P {Keith Watson}, The tensile strength of cohesive powders and its relationship to consolidation, free volume and cohesivity.
\newblock {\em\protect\JournalTitle{Powder Technology}} \textbf{97}, 237--245 (1998).

\bibitem{jackson00}
R Jackson, {\em The dynamics of fluidized particles}.
\newblock (Cambridge University Press), (2000).

\bibitem{Kranz2022}
WT Kranz, O Coquand, O D’Angelo, Understanding dense granular flow from first principles.
\newblock {\em\protect\JournalTitle{Sci. Talks}} \textbf{3}, 100049 (2022).

\bibitem{Krieger1959}
IM Krieger, TJ Dougherty, A mechanism for non-newtonian flow in suspensions of rigid spheres.
\newblock {\em\protect\JournalTitle{Trans. Soc. Rheol.}} \textbf{3}, 137--152 (1959).

\bibitem{Song2008}
C Song, P Wang, HA Makse, A phase diagram for jammed matter.
\newblock {\em\protect\JournalTitle{Nature}} \textbf{453}, 629--632 (2008).

\bibitem{abate+durian06}
AR Abate, DJ Durian, Approach to jamming in an air-fluidized granular bed.
\newblock {\em\protect\JournalTitle{Phys. Rev. E}} \textbf{74}, 031308 (2006).

\bibitem{Kranz2010}
WT Kranz, M Sperl, A Zippelius, Glass transition for driven granular fluids.
\newblock {\em\protect\JournalTitle{Physical Review Letters}} \textbf{104}, 225701 (2010).

\bibitem{Seguin2016}
A Seguin, O Dauchot, Experimental evidence of the gardner phase in a granular glass.
\newblock {\em\protect\JournalTitle{Phys. Rev. Lett.}} \textbf{117}, 228001 (2016).

\bibitem{Forterre2008}
Y Forterre, O Pouliquen, Flows of dense granular media.
\newblock {\em\protect\JournalTitle{Annual Review of Fluid Mechanics}} \textbf{40} (2008).

\bibitem{Madraki2020}
Y Madraki, et~al., Shear thickening in dense non-brownian suspensions: Viscous to inertial transition.
\newblock {\em\protect\JournalTitle{J. Rheol.}} \textbf{64}, 227--238 (2020).

\bibitem{Tapia2022}
F Tapia, M Ichihara, O Pouliquen, E Guazzelli, Viscous to inertial transition in dense granular suspension.
\newblock {\em\protect\JournalTitle{Phys. Rev. Lett.}} \textbf{129}, 078001 (2022).

\bibitem{Jenkins2007}
JT Jenkins, Dense inclined flows of inelastic spheres.
\newblock {\em\protect\JournalTitle{Granular Matter}} \textbf{10}, 47--52 (2007).

\bibitem{Spenley1996}
NA Spenley, XF Yuan, ME Cates, Nonmonotonic constitutive laws and the formation of shear-banded flows.
\newblock {\em\protect\JournalTitle{J. Phys. II}} \textbf{6}, 551--571 (1996).

\bibitem{Fielding2007}
SM Fielding, Complex dynamics of shear banded flows.
\newblock {\em\protect\JournalTitle{Soft Matter}} \textbf{3}, 1262--1279 (2007).

\bibitem{Stickel2005}
JJ Stickel, RL Powell, Fluid mechanics and rheology of dense suspensions.
\newblock {\em\protect\JournalTitle{Annual Review of Fluid Mechanics}} \textbf{37}, 129--149 (2005).

\bibitem{Barnes1987}
HA Barnes, MF Edwards, LV Woodcock, Applications of computer simulations to dense suspension rheology.
\newblock {\em\protect\JournalTitle{Chemical Engineering Science}} \textbf{42}, 591--608 (1987).

\bibitem{Fuchs2002}
M Fuchs, ME Cates, Theory of nonlinear rheology and yielding of dense colloidal suspensions.
\newblock {\em\protect\JournalTitle{Physical Review Letters}} \textbf{89}, 248304 (2002).

\bibitem{Brown2014}
E Brown, HM Jaeger, Shear thickening in concentrated suspensions: phenomenology, mechanisms and relations to jamming.
\newblock {\em\protect\JournalTitle{Reports on progress in physics}} \textbf{77} (2014).

\bibitem{vanMegen1995}
W van Megen, Crystallisation and the glass transition in suspensions of hard colloidal spheres.
\newblock {\em\protect\JournalTitle{Transp. Theory Stat. Phys.}} \textbf{24}, 1017--1051 (1995).

\bibitem{BiZhang2011}
D Bi, J Zhang, B Chakraborty, RP Behringer, Jamming by shear.
\newblock {\em\protect\JournalTitle{Nature}} \textbf{480}, 355--358 (2011).

\bibitem{Charbonneau2017}
P Charbonneau, J Kurchan, G Parisi, P Urbani, F Zamponi, Glass and jamming transitions: From exact results to finite-dimensional descriptions.
\newblock {\em\protect\JournalTitle{Annu. Rev. Condens. Matter Phys.}} \textbf{8}, 265--288 (2017).

\bibitem{Liu1998}
AJ Liu, SR Nagel, {Jamming is not just cool any more}.
\newblock {\em\protect\JournalTitle{Nature}} \textbf{396}, 21--22 (1998).

\bibitem{Gondret2002}
P Gondret, M Lance, L Petit, Bouncing motion of spherical particles in fluids.
\newblock {\em\protect\JournalTitle{Phys. Fluids}} \textbf{14}, 643--652 (2002).

\bibitem{Yang2008}
FL Yang, ML Hunt, A mixed contact model for an immersed collision between two solid surfaces.
\newblock {\em\protect\JournalTitle{Phil. Trans. Royal Soc. A}} \textbf{366}, 2205--2218 (2008).

\bibitem{Lorenz1997}
A Lorenz, C Tuozzolo, MY Louge, Measurements of impact properties of small, nearly spherical particles.
\newblock {\em\protect\JournalTitle{Exp. Mech.}} \textbf{37}, 292--298 (1997).

\bibitem{DeGiuli2016}
E DeGiuli, JN McElwaine, M Wyart, Phase diagram for inertial granular flows.
\newblock {\em\protect\JournalTitle{Phys. Rev. E}} \textbf{94}, 012904 (2016).

\bibitem{Biggs2008}
MJ Biggs, et~al., Granular temperature in a gas fluidized bed.
\newblock {\em\protect\JournalTitle{Granular Matter}} \textbf{10}, 63--73 (2008).

\bibitem{Cody1996}
GD Cody, DJ Goldfarb, GV Storch~Jr, AN Norris, Particle granular temperature in gas fluidized beds.
\newblock {\em\protect\JournalTitle{Powder Tech.}} \textbf{87}, 211--232 (1996).

\bibitem{Carnahan1969}
NF Carnahan, KE Starling, Equation of state for nonattracting rigid spheres.
\newblock {\em\protect\JournalTitle{J. Chem. Phys.}} \textbf{51}, 635--636 (1969).

\bibitem{Baker2018}
J Baker, F Guillard, B Marks, I Einav, {X}-ray rheography uncovers planar granular flows despite non-planar walls.
\newblock {\em\protect\JournalTitle{Nature Communications}} \textbf{9}, 5119 (2018).

\bibitem{Rosato2020}
A Rosato, K Windows-Yule, {Chapter 3} -- {I}nvestigative approaches {I}: {E}xperimental imaging techniques in {\em Segregation in Vibrated Granular Systems}, eds.{} A Rosato, K Windows-Yule.
\newblock (Academic Press), pp. 37--74 (2020).

\bibitem{Stannarius2017}
R Stannarius, Magnetic resonance imaging of granular materials.
\newblock {\em\protect\JournalTitle{Review of Scientific Instruments}} \textbf{88}, 051806 (2017).

\bibitem{Guo2021}
Q Guo, et~al., Dynamically structured bubbling in vibrated gas-fluidized granular materials.
\newblock {\em\protect\JournalTitle{Proceedings of the National Academy of Sciences}} \textbf{118}, e2108647118 (2021).

\bibitem{Taylor1923}
GI Taylor, Stability of a viscous liquid contained between two rotating cylinders.
\newblock {\em\protect\JournalTitle{Philosophical Transactions of the Royal Society of London}} \textbf{223}, 289--343 (1923).

\bibitem{Stuart1986}
JT Stuart, Taylor-vortex flow: a dynamical system.
\newblock {\em\protect\JournalTitle{SIAM Rev.}} \textbf{28}, 315--342 (1986).

\bibitem{Conway2004}
SL Conway, T Shinbrot, BJ Glasser, A taylor vortex analogy in granular flows.
\newblock {\em\protect\JournalTitle{Nature}} \textbf{431}, 433--437 (2004).

\bibitem{Krishnaraj2016}
KP Krishnaraj, PR Nott, A dilation-driven vortex flow in sheared granular materials explains a rheometric anomaly.
\newblock {\em\protect\JournalTitle{Nature Comm.}} \textbf{7}, 1--8 (2016).

\bibitem{Boyer2011}
F Boyer, {\'E} Guazzelli, O Pouliquen, Unifying suspension and granular rheology.
\newblock {\em\protect\JournalTitle{Phys. Rev. Lett.}} \textbf{107}, 188301 (2011).

\bibitem{Coquand2020}
O Coquand, M Sperl, WT Kranz, Integration through transients approach to the $\ensuremath{\mu}(\mathcal{I})$ rheology.
\newblock {\em\protect\JournalTitle{Physical Review E}} \textbf{102}, 032602 (2020).

\bibitem{Salehi2018}
H Salehi, D Sofia, D Sch{\"u}tz, D Barletta, M Poletto, Experiments and simulation of torque in anton paar powder cell.
\newblock {\em\protect\JournalTitle{Particulate Science and Technology}} \textbf{36}, 501--512 (2018).

\bibitem{Moller2009}
PCF M{\o}ller, A Fall, D Bonn, Origin of apparent viscosity in yield stress fluids below yielding.
\newblock {\em\protect\JournalTitle{Europhys. Lett.}} \textbf{87}, 38004 (2009).

\bibitem{Poeschel2001}
T Pöschel, S Luding, eds., {\em Granular Gases}.
\newblock (Springer-Verlag, Berlin Heidelberg), (2001).

\end{thebibliography}

\end{document}